\documentclass[aps, prb, superscriptaddress,amsmath,amssymb,twocolumn, longbibliography]{revtex4-2}
\usepackage{graphicx}
\usepackage{subfigure}
\usepackage{adjustbox}
\usepackage{bm}
\usepackage{color}
\usepackage{braket}
\usepackage{standalone}
\usepackage{multirow}
\usepackage{tikz}
\usepackage{mathrsfs}
\usepackage{dsfont}
\usepackage{comment}
\usepackage{amsmath}
\usepackage{amsmath,blkarray,booktabs}
\usepackage{float}
\usepackage{enumitem}
\usepackage[colorlinks,bookmarks=true,citecolor=blue,linkcolor=red,urlcolor=blue]{hyperref}
\usepackage{cleveref}
\usepackage{orcidlink}
\usepackage{appendix}
\usepackage{graphicx}
\usepackage{caption}
\usepackage{xcolor}
\usepackage[dvipsnames]{xcolor}
\usepackage{hhline}

\newcommand{\eps}{\varepsilon}
\newcommand{\bea}{\begin{eqnarray*}}
\newcommand{\eea}{\end{eqnarray*}}
\newcommand{\bean}{\begin{eqnarray}}
\newcommand{\eean}{\end{eqnarray}}
\newcommand{\lgl}{\langle}
\newcommand{\rgl}{\rangle}
\newcommand{\nn}{\nonumber\\}
\newcommand{\upa}{\uparrow}
\newcommand{\dwa}{\downarrow}
\renewcommand{\rm}{\mathrm}
\newcommand{\bs}{{\mathbf s}}
\newcommand{\br}{{\mathbf r}}

\newcommand{\bq}{{\mathbf q}}

\newcommand{\bA}{{\mathbf A}}

\newcommand{\bP}{{\mathbf P}}
\newcommand{\bR}{{\mathbf R}}
\newcommand{\bU}{{\mathbf U}}

\newcommand{\bu}{{\mathbf u}}

\newcommand{\btau}{\bm{\tau}}
\newcommand{\cM}{{\mathcal M}}

\newcommand{\cE}{{\mathcal E}}

\newcommand{\hc}{\hat c}

\newcommand{\hcd}{\hat c^\dagger}

\newcommand{\hpsi}{\hat \psi}
\newcommand{\hpsid}{\hat \psi^\dagger}

\newcommand{\bwe}{\begin{widetext}\begin{eqnarray}}
\newcommand{\eew}{\end{eqnarray}\end{widetext}}

\begin{document}

 \title{Skyrmion Excitations in the $\nu=-1$ Quantum Hall state in Monolayer Graphene }
\author{Jincheng An}
\email{jincheng.an1@gmail.com}
\affiliation{Shandong Provincial Key Laboratory of Light Field Manipulation Physics and Applications  \&
School of Physics and Optoelectronics, Shandong Normal University, Jinan 250358, China}

\author{Ajit C. Balram\orcidlink{0000-0002-8087-6015}}
\email{cb.ajit@gmail.com}
\affiliation{Institute of Mathematical Sciences, CIT Campus, Chennai 600113, India}
\affiliation{Homi Bhabha National Institute, Training School Complex, Anushakti Nagar, Mumbai 400094, India} 

\author{Ganpathy Murthy\orcidlink{0000-0001-8047-6241}}
\email{murthy@g.uky.edu}
\affiliation{Department of Physics and Astronomy, University of Kentucky, Lexington, KY 40506, USA}

\begin{abstract}
We investigate the skyrmion excited states atop the $SU(4)$ quantum Hall ferromagnetic ground state at filling factor $\nu=-1$ in monolayer graphene.  The competition among short-range anisotropic interactions, the Zeeman coupling, and sublattice symmetry-breaking potential gives rise to four distinct spin-valley-ordered ground-state phases. Combining effective field theory with the Hartree--Fock approximation, we develop a variational framework that incorporates both the long-range Coulomb interaction and the symmetry-breaking terms on an equal footing. Variational minimization determines the optimal skyrmion texture, including both its spatial radial profile and internal $SU(4)$ spinor structure. We establish the phase diagram of skyrmion excitations, identify thirteen distinct skyrmion phases under different external-field conditions, and determine their spin-valley textures, energies, and characteristic sizes. The robustness of the results against different radial ans\"atze demonstrates the reliability of our variational approach. Our work provides a systematic framework for studying skyrmion excitations in multicomponent quantum Hall ferromagnets and can be naturally extended to more general $SU(4)$ quantum Hall systems.
\end{abstract}

\maketitle

\section{Introduction}
Since the discoveries of the integer and fractional quantum Hall (QH) effects~\cite{IQHE_Discovery_1980, FQHE_Discovery_1982, Laughlin_1983}, two-dimensional electron systems in strong perpendicular magnetic fields have become a paradigmatic platform for studying topological phases of matter. QH states are the first discovered topological insulators, characterized by an incompressible bulk gap for charged excitations and gapless chiral edge modes that govern electrical and thermal transport.

The isolation of monolayer graphene in 2004~\cite{Berger_etal_2004, Novoselov_etal_2004, Zhang_etal_2005,neto2009:rmp} marked a major milestone for this field. Graphene consists of a single sheet of carbon atoms arranged in a honeycomb lattice with two sublattice sites (A and B) per unit cell. Owing to its remarkable properties, including ultrahigh carrier mobility, linear/Dirac dispersion near charge neutrality, and exceptional tunability, graphene has become the prototype of a broad family of atomically thin materials, including transition-metal dichalcogenides and various van der Waals heterostructures~\cite{Geim2013VanDW, Novoselov20162DMA}.

Over the past two decades, these systems have provided an unprecedented platform for exploring QH physics. The interplay of internal degrees of freedom, interaction-driven symmetry breaking, and topological order displayed by these materials has significantly advanced our understanding of both topological phases and correlated electron phenomena. Among them, graphene has played a particularly prominent role in revealing novel QH ferromagnetic phases~\cite{Shivaji_Skyrmion, QHFM_Yang_etal_1994, QHFM_Moon_etal_1995}, emergent collective excitations, and intricate patterns of spontaneous symmetry breaking.

Near charge neutrality, graphene’s low-energy electronic structure is governed by two inequivalent valleys, $K$ and $K'$, located at opposite corners of the Brillouin zone and related by time-reversal symmetry~\cite{neto2009:rmp}. In each valley, the carriers possess a linear, Dirac-like dispersion characteristic of massless relativistic fermions. When a perpendicular magnetic field $B_{\perp}$ is applied, this Dirac spectrum reorganizes into particle–hole symmetric Landau levels (LLs) labeled by $n=0,\pm1,\pm2,\ldots$, with energies scaling as $E_{\pm n}\propto\pm\sqrt{B_{\perp}|n|}$. A remarkable feature of the $n=0$ zeroth Landau level (ZLL) is its sublattice-valley locking: states in the two valleys reside exclusively on opposite sublattices of the honeycomb lattice~\cite{neto2009:rmp}. This locking plays a central role in the internal structure of QH states in the ZLLs of graphene.

Each LL manifold in graphene is (nearly) fourfold degenerate, reflecting the combination of two valley flavors and two spin states. Importantly, the Zeeman splitting $E_Z$ is typically the smallest energy scale. In the non-interacting limit and with $E_Z\rightarrow 0$, the Hamiltonian enjoys an approximate $SU(4)$ symmetry acting on the combined spin–valley space. The long-range part of the electron–electron Coulomb interaction, which dominates all other scales, also respects this symmetry. When a LL is partially filled, the Coulomb interaction drives a spontaneous breaking of the $SU(4)$ symmetry by aligning electrons in a particular spin-valley direction. This phenomenon is referred to as QH ferromagnetism (QHFM)~\cite{Shivaji_Skyrmion, QHFM_Yang_etal_1994, QHFM_Moon_etal_1995}. At charge neutrality ($\nu=0$), two of the four ZLL states are occupied, resulting in a highly degenerate $SU(4)$ manifold. While the long-range Coulomb interaction preserves this symmetry, the degeneracy is lifted by the Zeeman coupling $E_Z$, the sublattice symmetry-breaking potential $E_V$ induced by the hexagonal boron nitride (hBN) substrate~\cite{KV_DGG_2013, KV_expt_2013, KV_hBN_Jung2015, KV_hBN_Jung2017}, and short-range lattice-scale interactions~\cite{alicea2006:gqhe, KYang_SU4_Skyrmion_2006, Herbut1, Herbut2}.\\
Motivated by the discrepancy between the theoretical prediction of a quantum spin Hall state~\cite{abanin2006:nu0, Brey_Fertig_2006} and the experimentally observed insulating behavior at charge neutrality~\cite{young2014:nu0}, Kharitonov proposed a minimal interacting model for graphene in the ZLL~\cite{kharitonov2012:nu0}. Owing to lattice symmetries and momentum conservation~\cite{alicea2006:gqhe, KYang_SU4_Skyrmion_2006, Herbut1, Herbut2}, the short-range interactions preserve $SU(2)_\text{spin} \otimes U(1)_\text{valley}\otimes (Z_2)_\text{sublattice}$ symmetry and are completely parameterized by two anisotropy couplings, $u_z$ and $u_\perp$, corresponding to the valley-Ising and valley-$XY$ channels, respectively. Since these interactions originate from the lattice scale $a \ll \ell$, where $\ell=\sqrt{\hbar c/(eB_{\perp})}$ is the magnetic length, Kharitonov approximated them by ultra-short-range (USR) contact interactions. Within the Hartree-Fock approximation, this model yields four competing symmetry-broken QH ferromagnetic phases~\cite{kharitonov2012:nu0}—ferromagnetic (FM), canted antiferromagnetic (CAFM), bond-ordered (BO), and charge-density-wave (CDW)—which successfully account for the phase diagram observed at $\nu=0$~\cite{kharitonov2012:nu0}. Subsequent work has shown that including longer-range structure in the anisotopic interactions modifies Kharitonov's phase diagram by introducing new phases~\cite{Das_Kaul_Murthy_2022, Stefanidis_Sodemann2023}. Similar effects are observed for fractions as well~\cite{Jincheng2024, an2024fractional}.\\
Although this framework was originally developed for charge neutrality, the same symmetry constraints also apply at filling factor $\nu=-1$, at which only one of the four LLs in the ZLL is occupied. In this case, since there is only a single flavor of electron, USR interactions are inactive, and the finite-range structure of the anisotropic interactions must necessarily be taken into account. As shown in Refs.~\cite{Lian_Rosch_Goerbig_2016, Lian_Goerbig_2017}, relaxing the USR limit yields four distinct $\nu=-1$ QHFM ground-state phases. The phase of a given sample is determined by the signs of the longer-range pieces of $u_z$ and $u_\perp$. Thus, if one could experimentally distinguish the different phases, one would be able to obtain valuable information about the interactions. One way to distinguish phases is to look for their excitations. 

A hallmark of QHFMs is that charged excitations may no longer be the simple particle–like or hole-like quasiparticles of Hartree–Fock (HF) theory. Instead, the lowest-energy charged objects may be skyrmions~\cite{Shivaji_Skyrmion}, topologically nontrivial textures in the internal order-parameter space. In systems with only $SU(2)$ symmetry, such as traditional QHFM in two-dimensional electron gas (2DEG), skyrmions have been detected in the vicinity of $\nu=1$~\cite{Barrett_1995, Schmeller95, Aifer96} and $\nu=1/3$~\cite{Groshaus08, Balram15d}. The competition between Coulomb exchange (which favors a large, smoothly varying texture) and the Zeeman coupling (which favors a localized spin flip) determines the optimal skyrmion size and the number of spin reversals. This balance can be tuned by varying the ratio of these two energy scales and has been extensively studied using HF approaches~\cite{Fertig_skyrmion_1994, Fertig_skyrmion_1996, Fertig_skyrmion_1997} and exact diagonalization~\cite{He_PRB1996, MacDonald_PhysRevB_1996, MacDonald_PhysRevB_1998, Quinn_PhysRevB_2002, Jolicoeur_skymion_2019}.

In graphene, the approximate $SU(4)$ spin–valley symmetry endows skyrmions  with a substantially richer structure~\cite{skymion_Arovas_1999, Ezawa_prb2002, Ezawa_PhysRevB2005, Ezawa_PhysRevD2005, Smith_PhysRevB2007, Nomura_PhysRevB2008, Jain_2011MulticomponentFQ, Lian_Rosch_Goerbig_2016, Lian_Goerbig_2017, Goerbig_nu0_Skyrmion_Zoo_2021}. Skyrmions can now explore a much larger manifold of internal orientations, and their energetics and quantum numbers sensitively reflect how the $SU(4)$ symmetry is broken by interactions and external fields in the ground state. In graphene, skyrmions are expected to be the lowest-energy charged excitations at any integer filling except when all LLs of the topmost occupied LL manifold are full. 

In this work, we study the skyrmion excitations above the  QHFM ground states at filling factor $\nu=-1$ in monolayer graphene. At $\nu=-1$, in addition to the external one-body fields $E_Z$ (Zeeman) and $E_V$ (sublattice potential), the QHFM ground state is influenced by the odd relative angular momentum Haldane-pseudopotentials~\cite{Haldane_Pseudopot1983} of the short-range anisotropic interactions~\cite{Haldane_Pseudopot1983, Lian_Rosch_Goerbig_2016, Lian_Goerbig_2017}. The even relative angular momentum Haldane pseudopotentials do not enter, because the single-flavor forces spatial antisymmetry. The competition between the one-body terms and the anisotropic interactions gives rise to a ground-state phase diagram comprising four distinct QHFM phases~\cite{Lian_Rosch_Goerbig_2016, Lian_Goerbig_2017}. By combining effective field theory with the HF approximation, we incorporate the long-range Coulomb interaction, short-range anisotropic interactions, and external fields into a unified variational energy functional for the skyrmion order parameter. The minimization of this functional determines the optimal $SU(4)$ spinor profile of the skyrmions. A systematic scan over the interaction parameter space then yields the phase diagram of the competing skyrmion states.

Consistent with previous nomenclature, we will call the region in parameter space where a particular type of skyrmion has the lowest energy a skyrmion ``phase." Note that the same ground state phase can host multiple skyrmion phases. 

This system has been studied before~\cite{Lian_Rosch_Goerbig_2016, Lian_Goerbig_2017}. Our purpose in revisiting skyrmions at $\nu=-1$ is twofold: (i) To examine the effect of sublattice potentials on the skyrmions more carefully. It turns out that some of the skyrmion phases we find at nonzero $E_V$ were missed in previous work~\cite{Lian_Rosch_Goerbig_2016, Lian_Goerbig_2017}. Additionally, some skyrmion phases found to be distinct in previous work are found to be continuously connected in the presence a nonzero $E_V$. (ii) To propose and test more general ans\"atze for skyrmions in graphene, which we plan to use in the future at $\nu=0$. These ans\"atze remove infrared divergences in the variational energy, and allow us to analytically evaluate it. One of our findings is that the energy of the skyrmions is remarkably insensitive to the particular ansatz used. 

This work is organized as follows. In Sec.~\ref{sec:2}, we introduce the model Hamiltonian, which consists of the long-range Coulomb interaction, the short-range $SU(4)$ symmetry-breaking interactions, and anisotropic external fields. In Sec.~\ref{sec:3}, we treat the short-range anisotropic interactions within the HF approximation, derive the energy functional for the QHFM ground states, and reproduce the $\nu=-1$ phase diagram obtained previously~\cite{Lian_Rosch_Goerbig_2016, Lian_Goerbig_2017}. In Sec.~\ref{sec:4}, we combine effective field theory for the long-range Coulomb interaction with the HF approximation for the short-range anisotropic interactions to construct a variational energy functional for skyrmion excitations. In Sec.~\ref{sec:5}, we present the resulting skyrmion phase diagrams and discuss the spinor structures, energies, and optimal sizes of the various skyrmion states. In Sec.~\ref{sec:6}, we compare the results obtained from several representative variational ans\"aze, thereby confirming the robustness of our variational approach. Finally, we summarize our main results and provide an outlook in Sec.~\ref{sec:conclusion}. Mathematical details are relegated to the appendices. 

\section{The Hamiltonian of Monolayer Graphene}\label{sec:2}
The interacting part of the Hamiltonian for monolayer graphene within the ZLL is~\cite{kharitonov2012:nu0} 
\bean
\hat H=\hat H_{\rm {Coul}}+\hat H_\text{an},
\eean 
where 
\bean\label{eq:long_range_coulomb}
&&\hat H_{\rm {Coul}}=\dfrac{1}{2}\dfrac{e^2}{\eps}\int d^2\br_1d^2\br_2\nn
&&\quad\quad\quad:\hpsid(\br_1)\hpsi(\br_1)\dfrac{1}{|\br_1-\br_2|}\hpsid(\br_2)\hpsi(\br_2):
\eean 
is the $SU(4)$ invariant long-range Coulomb interaction. More importantly, $\hat H_\text{an}$  contains the lattice-scale residual anisotropic interaction which, in conjunction with the one-body couplings, will determine the orientation of the spin-valley QHFM. Parameterized by two independent radial functions $V_z(r)$ and $V_x(r)=V_y(r)=V_\perp(r)$, the anisotropic interactions take the following form 
\bean\label{eq: anisotropic_interaction}
\hspace{-0.5cm}&&\hat H_\text{an}=\dfrac{1}{2}\sum_{a=x,y,z}\int d^2\br_1d^2\br_2:\hat\btau^a(\br_1)V_a(\br_1-\br_2)\hat\btau^a(\br_2):,\nn
\hspace{-0.5cm}&&\quad\hat\btau^a(\br_1)=\sum_{\alpha\gamma}\hpsid_\alpha(\br_1) \tau^{a}_{\alpha\gamma}\hpsi_\gamma(\br_1),
\eean
where $\tau^{a}\equiv\tau^a_\text{valley}\otimes\sigma^0_\text{spin}$, with $a=x,y,z$, are the direct products of Pauli matrices acting on the four-dimensional spin/valley space and $\alpha,\ \gamma=K\upa,\ K\dwa,\ K^\prime\upa,\ K^\prime\dwa$. After expanding the creation and destruction operators in terms of the ZLL eigenstates of the non-interacting Hamiltonian in the symmetric gauge 
\bean\label{eq:field_operator_expansion}
\hpsi_\alpha(\br)=\sum_m\phi_m(\br)\hc_{m,\alpha},\ 
\phi_m(z)=\dfrac{(z/\ell)^m}{\sqrt{2\pi 2^m m!}\ell} e^{-\frac{z\bar z}{4\ell^2}},
\eean 
the interaction can be expressed as 
\bean
&&\hat H_\text{an}=\dfrac{1}{2}\sum_aV^a_{m_1m_2m_3m_4}:\hat\tau^a_{m_1m_4}\hat\tau^a_{m_2m_3}:,\nn
&&V^a_{m_1m_2m_3m_4}=\lgl m_1,m_2|V_a(\br_1-\br_2)|m_3,m_4\rgl,\nn
&&\hat\tau^a_{m_1m_4}=\sum_{\alpha\gamma}\hcd_{m_1,\alpha}\tau^a_{\alpha\gamma}\hc_{m_4,\gamma}.
\eean
We Fourier transform the interaction and expand in terms of Haldane pseudopotentials~\cite{Haldane_Pseudopot1983} as
\bean\label{eq:vaq_def}
&&V_a(\br_1-\br_2)=\frac{1}{(2\pi)^2}\int d^2\bq~e^{-i \bq\cdot(\br_1-\br_2)}v_a(\bq),\nn
&&v_a(\bq)=\sum_m 4\pi\ell^2u^{(m)}_aL_m(q^2\ell^2).
\eean
Here $u^{(m)}_a$ are the Haldane pseudopotentials in the relative angular momentum channel $m$ and $L_m$ are Laguerre polynomials. 

The interaction matrix elements $V^a_{m_1m_2m_3m_4}$ can be decomposed in terms of center-of-mass and relative angular momenta as~\cite{Sodemann_MacDonald_LLmix}
\bean
\hspace{-0.5cm}&&V^a_{m_1m_2m_3m_4}=\sum_mu^{(m)}_a\bU^{(m)}_{m_1m_2m_3m_4},\nn
\hspace{-0.5cm}&&\bU^{(m)}_{m_1m_2m_3m_4}=\sum_M \langle m_1,m_2|M,m\rangle\langle M,m|m_3,m_4\rangle.
\eean 
For illustration, the coefficients that combine two individual angular momenta $m_{1}$ and $m_{2}$ into the center-of-mass $M$ and relative $m$ angular momenta for $m=0,1$ are~\cite{Sodemann_MacDonald_LLmix}
\begin{eqnarray}
\hspace{-0.2cm}&\langle m,m'\lvert m+m',0\rangle=\dfrac{1}{\sqrt{2^{m+m'}}}\sqrt{\dfrac{(m+m')!}{m!m'!}}\nonumber\\
\hspace{-0.2cm}&\langle m,m'\lvert m+m'-1,1\rangle=\dfrac{m-m'}{\sqrt{2^{m+m'}}}\sqrt{\dfrac{(m+m'-1)!}{m!m'!}}.
\end{eqnarray}
Next, we include the Zeeman and valley Zeeman terms in the Hamiltonian.
\bean\label{eq:Zeeman_term}
&&\hat H_Z=-E_Z\sum_m\hat\sigma_m^z,\quad \hat H_V=-E_V\sum_m\hat\tau_m^z,\nn
&&\hat\sigma_m^z=\sum_{\alpha\gamma}\hat c^\dagger_{m,\alpha}(\tau^0_\text{valley}\otimes\sigma_\text{spin}^z)_{\alpha\gamma}\hat c_{m,\gamma},\nn
&&\hat\tau_m^z=\sum_{\alpha\gamma}\hat c^\dagger_{m,\alpha}(\tau^z_\text{valley}\otimes\sigma_\text{spin}^0)_{\alpha\gamma}\hat c_{m,\gamma}.
\eean 
Throughout this work, we choose to fix the magnetic field at $B_{\perp}{=}10~\text{Tesla}$, implying a Zeeman energy of $e_Z{=}(1/2)g_{B}\mu_B B_{\perp}{=}0.58~\text{meV}$ [the $g$-factor $g_{B}{=}2$ for graphene]. In later phase diagrams, we will take the $e_Z$ as the energy unit, i.e., $e_Z{=}1$. The entire Hamiltonian is
\begin{equation}
    \hat H=\hat H_{\rm {Coul}}+\hat H_\text{an}+\hat H_Z+\hat H_V
\end{equation}

\section{Quantum Hall Ferromagnetic Ground States}\label{sec:3}

We assume that the $\nu=-1$ ground state $|\Phi\rgl$ is a single Slater determinant (SSD) with translational invariance up to intervalley coherence. Any SSD is uniquely specified by its one-body averages,
\bean
\lgl\Phi|\hcd_{m,\beta}\hc_{m^\prime,\eta}|\Phi\rgl=\delta_{m,m^\prime}\Delta_{\eta\beta},
\eean
The matrix $\Delta$ is simply the projector to the occupied subspace in the four-dimensional spin/valley space. 
Since the Coulomb interaction is ineffective in selecting the ground state, we will ignore it in this section. Using the HF approximation, we obtain the ground state energy functional 
\bean\label{eq:ground_state_energy}
\hspace{-0.5cm}E_\text{g.s.}[\Delta]&=&\dfrac{1}{N_\phi}\lgl\Phi|\hat H_\text{an}+\hat H_Z|\Phi\rgl\nn
\hspace{-0.5cm}&=&\dfrac{1}{2}\sum_a\Big[u_{a,H}\text{Tr}\big(\tau^a\Delta\big)^2-u_{a,F}\text{Tr}\big(\tau^a\Delta\tau^a\Delta\big)\Big]\nn
\hspace{-0.5cm}&&\ -E_Z\text{Tr}\big(\sigma^z\Delta\big)-E_V\text{Tr}\big(\tau^z\Delta\big),
\eean 
with the Hartree and Fock couplings~\cite{an2025_transport_gap} 
\bean\label{eq:HF_couplings}
\hspace{-0.5cm}&&u_{a,H}=\dfrac{1}{N_\phi}\sum_{m_1,m_2}V^a_{m_1m_2m_2m_1}=2\sum_m u^{(m)}_a,\nn
\hspace{-0.5cm}&&u_{a,F}=\dfrac{1}{N_\phi}\sum_{m_1,m_2}V^a_{m_1m_2m_1m_2}=2\sum_m(-1)^mu_a^{(m)},
\eean 
where we have used the following properties of the interaction matrix elements $\bU^{(m)}_{m_1m_2m_3m_4}$~\cite{an2024fractional, Jincheng2024, an2025_transport_gap}, 
\bean\label{sm_eq_UU}
{\mathbf U}^{(m)}_{m_1m_2m_2m_1}&=&(-1)^m{\mathbf U}^{(m)}_{m_1m_2m_1m_2},\nn
\sum_{m_1, m_2}{\mathbf U}^{(m)}_{m_1m_2m_2m_1}&=&2N_\phi.
\eean
In the USR limit~\cite{kharitonov2012:nu0}, $V_a(\br_1-\br_2)\propto\delta(\br_1-\br_2)$, and only the lowest Haldane pseudopotential $u^{(0)}_a$ is nonzero, thus, we will have $u_{a, H}=u_{a, F}$ by Eq.~\eqref{eq:HF_couplings} while for a generic interaction, the parameter space for the anisotropic interaction is four-dimensional $(u_{\perp, H},\ u_{\perp, F},\ u_{z, H},\ u_{z, F})$.

At filling $\nu=-1$, we need to select a particular linear combination of the four basis  spinors to be occupied. In this situation, the following ansatz for the four spinors furnishes a complete parametrization of the mean-field ground-state manifold, capturing all possible competing ordered states.

\begin{figure}[h]
    \centering \includegraphics[width=0.49\textwidth,height=0.49\textwidth]{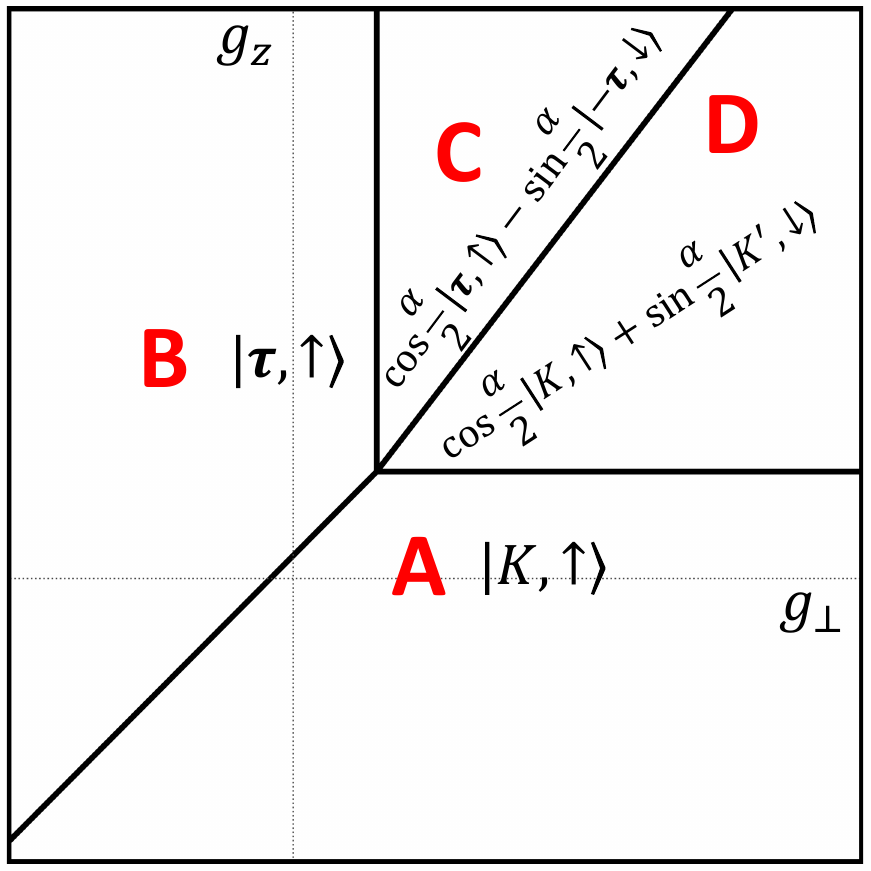}
    \caption{The $\nu=-1$ QHFM ground state phase diagram with $ g_{\perp},g_{z}\in [-400,800]~\text{meV}\cdot \text{nm}^{2}$. External fields are fixed at $E_Z=1e_Z\ \&\ E_V=0.3e_Z$. There are four different phases: A, B, C, \&\ D. The occupied spinor for a given phase is presented in the region occupied by the phase.
    }
    \label{fig_ground_state_phase_diagram}
\end{figure}

\bean\label{eq:f_ansatz}
&&\lvert f_1\rangle =\cos\dfrac{\alpha_1}{2}\lvert\btau,\bs_a\rangle+\sin\dfrac{\alpha_1}{2}\lvert-\btau,-\bs_b\rangle,\nn
&&\lvert f_2\rangle =\cos\dfrac{\alpha_2}{2}\lvert\btau,-\bs_a\rangle+\sin\dfrac{\alpha_2}{2}\lvert-\btau,\bs_b\rangle,\nn
&&\lvert f_3\rangle =\sin\dfrac{\alpha_1}{2}\lvert\btau,\bs_a\rangle-\cos\dfrac{\alpha_1}{2}\lvert-\btau,-\bs_b\rangle,\nn
&&\lvert f_4\rangle =\sin\dfrac{\alpha_2}{2}\lvert\btau,-\bs_a\rangle-\cos\dfrac{\alpha_2}{2}\lvert-\btau,\bs_b\rangle\nn
&&\lvert\btau\rangle=\begin{pmatrix}
    \cos\frac{\theta_{\tau}}{2}\\
    e^{i\phi_{\tau}}\sin\frac{\theta_{\tau}}{2}
\end{pmatrix},\quad
  \lvert\bs\rangle=\begin{pmatrix}
    \cos\frac{\theta_{s}}{2}\\
    e^{i\phi_{s}}\sin\frac{\theta_{s}}{2}
\end{pmatrix},
\eean
where $|\btau,\bs\rangle {\equiv} |\btau\rangle{\otimes}|\bs\rangle$ with $\btau$ ($\bs$) being unit vectors on the valley (spin) Bloch spheres. The $U(1)_\text{valley}$ symmetry permits the choice $\phi_{\tau}=0,\pi$, which means we are working with real valley spinors. Likewise, the $U(1)_\text{spin}$ symmetry of spin rotations around the direction of total field allows us to choose $\phi_{s}=0,\pi$, which means we can work with real spin spinors. Thus, all four four-dimensional spinors can be chosen to be real.
The projector for filling $\nu=-1$ can be constructed as 
\bean
\Delta=|f_1\rgl\lgl f_1|.
\eean 
With $\text{Tr}\big(\tau^a\Delta\tau^a\Delta\big)=\text{Tr}\big(\tau^a\Delta\big)^2=\lgl f_1|\tau^a|f_1\rgl^2\equiv \big(\tau^a_{11}\big)^2$, the ground state energy functional of Eq.~\eqref{eq:ground_state_energy} will simplify to 
\bean\label{eq:_m1_ground_state_energy}
E_\text{g.s.}=\dfrac{1}{2}\sum_a\big(u_{a,H}-u_{a,F}\big)\big(\tau^a_{11}\big)^2-E_Z\sigma^z_{11}-E_V\tau^z_{11},
\eean
which, by Eq.~\eqref{eq:HF_couplings}, indicates that $\nu=-1$ QHFM state only responds to the interacting channels $u^{(m)}_a$ with odd $m$. Physically, this is because the single flavor forces spatial antiymmetry between the electrons, which can only occur for $m$ odd. Thus, the minimal construction of the anisotropic interaction for the $\nu=-1$ QHFM ground states keeps the lowest odd Haldane pseudopotential $u_a^{(1)}\neq0,\ u^{(m>1)}_a=0$~\cite{Goerbig_nu0_Skyrmion_Zoo_2021}. In this case, we have
\bean\label{eq:viq}
v_a(\bq)=g_aL_1(q^2\ell^2).
\eean 
The resulting Haldane pseudopotentials are $u^{(1)}_a=g_a/(4\pi\ell^2)$,
yielding the Hartree and Fock couplings
\bean\label{eq:u1_HF_couplings}
u_{\perp,H}=-u_{\perp,F}=\dfrac{g_\perp}{2\pi\ell^2},\quad 
u_{z,H}=-u_{z,F}=\dfrac{g_z}{2\pi\ell^2}.
\eean 
As the ground state energy functional $E_\text{g.s.}$ in Eq.~\eqref{eq:_m1_ground_state_energy} only depends on the differences between the Hartree and Fock couplings, i.e., $u_{a, H}-u_{a, F}=g_a/(\pi\ell^2)$, minimizing $E_\text{g.s.}$ over the anisotropic interaction parameter space $(g_\perp,g_z)$ completely determines the ground-state phase diagram.\\
The $\nu=-1$ QHFM ground state phase diagram is presented in Fig.~\ref{fig_ground_state_phase_diagram}, which consists of four different phases labeled by A, B, C, \& D~\cite{Lian_Rosch_Goerbig_2016, Lian_Goerbig_2017}. The filled spinors in each phase are presented below.

\begin{itemize}[left=0pt]
\item In phase A, the occupied spinor is 
\bean\label{eq:phase_A_spinor}
&&|f_1^\text{A}\rgl=|K,\upa\rgl,\nn
&&E^\text{A}_\text{g.s}=\dfrac{1}{2}\Big(u_{z,H}-u_{z,F}-2E_Z-2E_V\Big).
\eean 

\item In phase B, the occupied spinor is 
\bean\label{eq:phase_B_spinor}
\hspace{-1cm}&&|f_1^\text{B}\rgl=|\btau,\upa\rgl,\quad |\btau\rgl=\begin{pmatrix}
    \cos\frac{\theta_{\tau}}{2}\\
    \sin\frac{\theta_{\tau}}{2}
\end{pmatrix},\nn
\hspace{-1cm}&&\cos\theta_\tau=\dfrac{E_V}{u_{z,H}-u_{z,F}-u_{\perp,H}+u_{\perp,F}},\nn
\hspace{-1cm}&&E^\text{B}_\text{g.s.}=\dfrac{1}{2}\Big(u_{\perp,H}-u_{\perp,F}-2E_Z-E_V\cos\theta_\tau\Big),
\eean 

\item In phase C, the occupied spinor is 
\bean\label{eq:phase_C_spinor}
\hspace{-1cm}&&|f_1^\text{C}\rgl=\cos\dfrac{\alpha}{2}|\btau,\upa\rgl-\sin\dfrac{\alpha}{2}|-\btau,\dwa\rgl,\nn
\hspace{-1cm}&&\cos\alpha=\dfrac{E_Z}{u_{\perp,H}-u_{\perp,F}},\nn
\hspace{-1cm}&&\cos\theta_\tau=\dfrac{E_V\big(u_{\perp,H}-u_{\perp,F}\big)}{E_Z\big(u_{z,H}-u_{z,F}-u_{\perp,H}+u_{\perp,F}\big)},\nn
\hspace{-1cm}&&E^\text{C}_\text{g.s.}=\dfrac{E_Z^2}{2\big(u_{\perp,H}-u_{\perp,F}\big)}-\dfrac{E_VE_Z\cos\theta_\tau}{2\big(u_{\perp,H}-u_{\perp,F}\big)}.
\eean 
\item In phase D, the occupied spinor is 
\bean\label{eq:phase_D_spinor}
\hspace{-1cm}&&|f_1^\text{D}\rgl=\cos\dfrac{\alpha}{2}|K,\upa\rgl+\sin\dfrac{\alpha}{2}|K^\prime,\dwa\rgl,\nn
\hspace{-1cm}&&\cos\alpha=\dfrac{E_Z+E_V}{u_{z,H}-u_{z,F}},\quad E^\text{D}_\text{g.s.}=\dfrac{\big(E_Z-E_V\big)^2}{2\big(u_{z,F}-u_{z,H}\big)}.
\eean 
\end{itemize}
It is worth noting that in the limit of  $E_V\rightarrow 0$, the valley Bloch vector $\btau$ in phases B \& C will simplify to 
$|\btau\rgl=\dfrac{1}{\sqrt 2}\big(|\upa\rgl+|\dwa\rgl\big)\equiv |\hat e_x\rgl,\ |-\btau\rgl=\dfrac{1}{\sqrt 2}\big(|\upa\rgl-|\dwa\rgl\big)\equiv |-\hat e_x\rgl$.

\section{Skyrmion Excitations of $\nu=-1$ QHFM}\label{sec:4}
\begin{figure}[H]
    \centering \includegraphics[width=0.49\textwidth,height=0.18\textwidth]{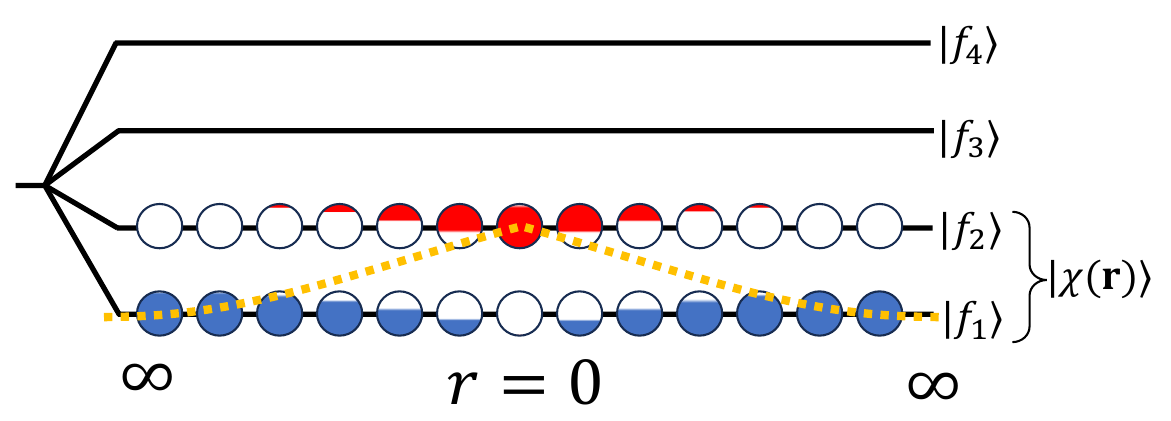}
    \caption{Schematic representation of a skyrmion excited state upon the $\nu=-1$ QHFM ground state. The skyrmion spinor $|\chi(\br)\rgl$ is a linear combination between the occupied spinor $|f_1\rgl$ and an orthogonal spinor  $|f_2\rgl$, which is unoccupied in the ground state, but occupied at the core of the skyrmion.
    }
    \label{fig_skyrmion_cartoon}
\end{figure}

Now we are ready to study the skyrmion excited states variationally. We start with the following general ansatz for a skyrmion above the $\nu=-1$ ground state, the projector for which is given by:
\bean\label{eq:P(r)}
\hspace{-1cm}&&P^S(\br)=|\chi(\br)\rgl\lgl \chi(\br)|,\nn
\hspace{-1cm}&&|\chi(\br)\rgl=\sqrt{1-W(r)}\dfrac{x+\text{i}y}{r}|f_1\rgl+\sqrt{W(r)}|f_2\rgl,
\eean
with $W(r)$ being an arbitrary smooth radial profile satisfying 
\bean\label{eq:W_conditions}
0\leq W(r)\leq 1,\quad W(0)=1,\quad W(\infty)=0.
\eean 
This ansatz makes sure that the skyrmion reverts to ground state spinor $|f_1\rgl$ as $r\to\infty$. A schematic representation of this skyrmion state is shown in Fig.~\ref{fig_skyrmion_cartoon}. Note that the variational parameters are the components of the spinor $|f_2\rgl$ occupied at the center, which has to be orthogonal to $|f_1\rgl$, and the function $W(r)$. 

The nonzero matrix elements of $P^S(\br)$ in the basis $|f_1\rgl,\ |f_2\rgl$ are given by
\bean\label{eq:skyr_p_matrix_elements}
\hspace{-1cm}&&P^S_{11}(\br)=1-W(r),\quad P^S_{22}(\br)=W(r),\nn
\hspace{-1cm}&&P^S_{12}(\br)=P^{S*}_{21}(\br)=\sqrt{1-W(r)}\sqrt{W(r)}\dfrac{x+\text{i}y}{r}.
\eean 
The topological charge density is  
\bean\label{eq:topological_charge_density}
{\varrho}(\br)=\dfrac{1}{2\pi i}\varepsilon_{\mu\nu}\text{Tr}\big[P^S\partial_\mu P^S\partial_\nu P^S\big]=-\dfrac{W^\prime(r)}{2\pi r},
\eean
leading to the topological charge 
\bean\label{eq:topo_charge}
Q=\int d^2\br{\varrho}(\br)=-W(r)|_0^\infty=1,
\eean
as required for a unit-charge skyrmion.\\

We will evaluate the energy of the skyrmion using the nonlinear sigma model (NLSM), which is a long-wavelength effective theory of QHFM~\cite{Shivaji_Skyrmion}. The total variational energy in this approach is given by 
\begin{equation}
    E=E_{\rm {stiff}}+E_{\rm {Coul}}+E_{\rm {SB}},
\end{equation}
where $E_{\rm {stiff}}$ is the spin/valley stiffness energy which penalizes changes in the order parameter, $E_{\rm {Coul}}$ is the Coulomb energy, and $E_{\rm {SB}}$ is the energy of symmetry-breaking terms including the one-body fields and the short-range anisotropic interactions. All these contributions turn out to be functionals of the NLSM order parameter $\bP(\br)$, which is simply the projector to the occupied state at $\br$.

Since $E_{\rm {stiff}}$ and $E_{\rm {Coul}}$ are typically much larger than $E_{\rm {an}}$, let us evaluate them first. 
\bean
\hspace{-0.9cm}&&E_{\rm {stiff}}=\rho_s\int d^2\br\text{Tr}\big[\partial_\mu \bP\partial_\mu \bP\big],\label{eq:NLSM}\\
\hspace{-0.9cm}&&E_{\rm{Coul}}=\dfrac{1}{2}\dfrac{e^2}{\eps}\int d^2\br_1d^2\br_2\varrho(\br_1)\dfrac{1}{|\br_1-\br_2|}\varrho(\br_2).
\eean 
where $\rho_s=1/(16\sqrt{2\pi})~e^2/(\varepsilon\ell)$ is the spin stiffness~\cite{QHFM_Moon_etal_1995}  with $\varepsilon\approx6$ being the dielectric constant of encapsulated graphene~\cite{Hunt2016DirectMO}.\\

Let us define the shape of the skyrmion that minimizes the stiffness energy to be $W_{0}(r)$. The calculation of $W_0(r)$ is outlined in Appendix~\ref{appdx_1}, where it is shown that~\cite{Bogomolny:1975de,Prasad_Sommerfield_1975,Goerbig_nu0_Skyrmion_Zoo_2021, Lian_Goerbig_2017, Macfarlane1979} 
\bean\label{eq:w_bps}
W_0(r)=\dfrac{\lambda^2}{r^2+\lambda^2},
\eean 
where $\lambda$ is the characteristic skyrmion size. This profile saturates the Bogomol\'nyi–Prasad–Sommerfield (BPS) bound~\cite{Bogomolny:1975de,Prasad_Sommerfield_1975} and yields the minimum NLSM stiffness energy, $E_{\rm {stiff}}=4\pi\rho_s$. \\

\begin{figure}[H]
    \centering \includegraphics[width=0.45\textwidth,height=0.35\textwidth]{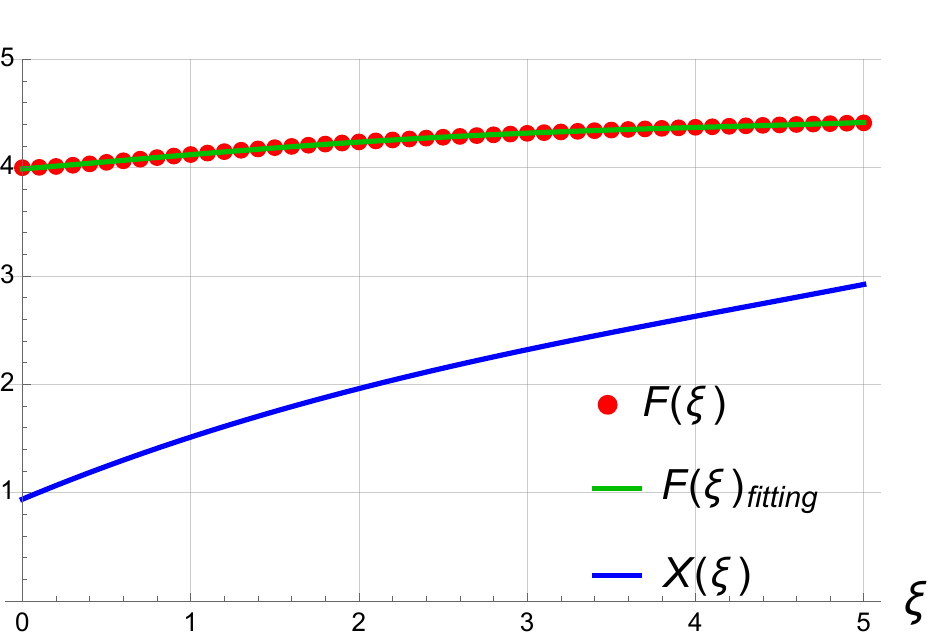}
    \caption{Dependence of the dimensionless functions $F(\xi)$ and $X(\xi)$ on $\xi$, defined in Eqs.~\eqref{eq:nlsm_energy} and \eqref{eq:Ec}, respectively. The monotonic increase of both functions indicates that increasing $\xi$ raises both the stiffness energy of the nonlinear sigma model and the Coulomb interaction energy of the skyrmion.
    }
    \label{fig_ft_X_xi}
\end{figure}

Due to the slow falloff of $W(r)$ with $r$, some of the terms in the one-body and anisotropic energies are divergent in the infrared. For example, it is evident by inspection that the one-body contributions to $E_{\rm {an}}$ are divergent in the size of the system. To remove this feature, we introduce a one-parameter family of ans\"atze inspired by $W_0(r)$,
\bean\label{eq:w(r)}
W_\xi(r)=\bigg(\dfrac{\lambda^2}{\lambda^2+r^2}\bigg)^{1+\xi},
\eean 
where $\xi\ge0$ is a dimensionless variational parameter characterizing the deviation of our ansatz from the BPS solution. For any $\xi>0$, all the terms in the variational energy converge, whereas, by construction,  at $\xi=0$ we have $W_{\xi=0}(r)=W_0(r)$. The corresponding topological charge density is Eq.~\eqref{eq:topological_charge_density}, becomes
\bean\label{eq:topo_density_xi}
\varrho_\xi(\br)=\dfrac{1+\xi}{\pi\lambda^2}\bigg(\dfrac{\lambda^2}{\lambda^2+r^2}\bigg)^{2+\xi}.
\eean 
In what follows, we will drop the subscript $\xi$ from $W_\xi(r)$ and $\varrho_\xi(\br)$ for brevity of notation. 
Substituting the order parameter in Eq.~\eqref{eq:P(r)} into Eq.~\eqref{eq:NLSM}, the stiffness energy is
\bean\label{eq:nlsm_energy}
&&\quad E_\text{stiff}\nn
&&=\rho_s\int d^2\br\bigg[\dfrac{2W(r)\Big(1-W(r)\Big)}{r^2}+\dfrac{W^\prime(r)^2}{2W(r)\Big(1-W(r)\Big)}\bigg]\nn
&&=\pi\rho_s\int_0^\infty du\dfrac{2}{u}\left(\frac{1}{u+1}\right)^{\xi +2} \nn
&&\quad\quad\times\left(\frac{(\xi +1)^2 u^2}{u+1-\left(\frac{1}{u+1}\right)^{\xi }}+u+1-\left(\frac{1}{u+1}\right)^{\xi }\right)\nn
&&\equiv \pi\rho_s F(\xi),
\eean 
where, in the second equality, we have substituted the variational ansatz in Eq.~\eqref{eq:w(r)}, introduced the variable $u=r^2$, and set $\lambda=1$ since this integral is independent of the characteristic length $\lambda$. The integral $F(\xi)$ can be evaluated analytically only for integer values of $\xi$,  yielding
\bean
&&F(0)=4,\quad F(1)=29/3-8\log 2,\nn
&&F(2)=587/30-\sqrt{3}\pi-9\log3,\nn
&&F(3)=7039/210-4\pi-24\log2.
\label{eq:F for integer xi}\eean 
For general $\xi$, $F(\xi)$ is evaluated numerically and is accurately approximated by the polynomial fit
\bean\label{eq:nlsm_fitting}
&&F(\xi)=4+0.129109\xi+0.01007\xi^2 \nn
&&\quad\quad\quad\quad- 0.008012\xi^3+0.000853\xi^4,
\eean 
which, as shown in Fig.~\ref{fig_ft_X_xi}, is monotonically increasing with $\xi$ in the range we consider.\\
To evaluate the Coulomb self-energy $E_\text{Coul}$, we first Fourier transform the topological charge density,
\bean
\tilde\varrho(\bq)=\int d^2\br\varrho(\br)e^{i\bq\cdot\br}
=\dfrac{(q\lambda)^{1+\xi}K_{1+\xi}(q\lambda)}{2^\xi\Gamma(1+\xi)},
\eean 
which satisfies $\tilde\varrho(-\bq)=\tilde\varrho(\bq)$. Then we have
\bean\label{eq:Ec}
&&E_\text{Coul}=\dfrac{1}{2}\dfrac{e^2}{\eps}\int_0^\infty dq\tilde\varrho(\bq)\tilde\varrho(-\bq)=\dfrac{1}{2}\dfrac{e^2}{\eps\lambda}X(\xi),\nn
&&X(\xi)=\dfrac{\pi\Gamma(3/2+\xi)\Gamma(5/2+2\xi)}{4^{1+\xi}\Gamma^2(1+\xi)\Gamma(2+\xi)}.
\eean 
The dimensionless function $X(\xi)$ increases monotonically with $\xi$, as shown in Fig.~\ref{fig_ft_X_xi}, and satisfies $X(0)=3\pi^2/32$.

We finally turn to the evaluation of $E_\text{SB}$ in the context of the NLSM. Recall that 
since the long-range Coulomb interaction is $SU(4)$ invariant, it is unable to select the optimal occupied spinor at the center of the skrmion $|\chi(0)\rgl=|f_2\rgl$. We need to compute
\bean
E_\text{SB}=\lgl\Psi_s|\hat H_\text{an}+\hat H_Z+\hat H_V|\Psi_s\rgl.
\eean 
Furthermore, we need to express this as a functional of $\bP(\br)$. This can be done, subject to the following conditions: (i) The scale of variation of the skyrmion should be much larger than the magnetic length, i.e., $\lambda\gg\ell$, and (ii) The anisotropic interactions should be small $u_a\left(\frac{\ell}{\lambda}\right)^2\ll\rho_s$. The details are presented 
in Appendix.~\ref{appdx_2}. We reproduce the relevant equations from Appendix~~\ref{appdx_2} [Eq.~\eqref{eq:app_symmetry_breaking_energy}] here:
\bean
E_\text{SB}=\dfrac{A}{2\pi\ell^2}E_\text{g.s.}+\dfrac{\lambda^2}{\ell^2}\bigg[\dfrac{\cE_1}{2\xi}+\dfrac{\cE_2}{2(1+2\xi)}\bigg].
\eean 
Note that $A$ is the area of the system. The first term is divergent but harmless, since it represents the ground state energy, which will be subtracted from the total energy to find the excitation energy of the skyrmion.   $\cE_1,\ \cE_2$ are functionals of $|f_2\rgl$, their explicit forms are derived in Appendix~\ref{appdx_2} [see Eqs.~\eqref{eq:app_E1} and \eqref{eq:app_E2}], and reproduced here.
\bean
\hspace{-1cm}&&\cE_1=\big(u_{a,H}-u_{a,F}\big)\Big[\tau^a_{12}\tau^a_{21}+\tau^a_{11}\tau^a_{22}-\big(\tau^a_{11}\big)^2\Big]\nn
\hspace{-1cm}&&\quad\quad\ +E_Z\big(\sigma^z_{11}-\sigma^z_{22}\big)+E_V\big(\tau^z_{11}-\tau^z_{22}\big),\label{eq:skyr_E1}\\
\hspace{-1cm}&&\cE_2=\dfrac{1}{2}\big(u_{a,H}-u_{a,F}\big)\Big[\big(\tau^a_{11}-\tau^a_{22}\big)^2-2\tau^a_{12}\tau^a_{21}\Big]\label{eq:skyr_E2},
\eean 
where $\tau^a_{ij}=\langle f_i|\tau^a|f_j\rangle$. Note that the appearance of $u_{a,H}-u_{a,F}$ as prefactors in these expressions shows that the even angular momentum Haldane pseudopotentials do not affect the energy of the skyrmions. 

After collecting all the contributions, we find the energy of the variational skyrmion.
\bean\label{eq:skyrmion_varitational_energy}
E_\text{skyr}=\pi\rho_sF(\xi)+\dfrac{e^2X(\xi)}{2\epsilon\lambda}+\dfrac{\lambda^2}{\ell^2}\bigg[\dfrac{\cE_1}{2\xi}+\dfrac{\cE_2}{2(1+2\xi)}\bigg],
\eean 
The excitation energy of the skyrmion has no infrared divergence for $\xi>0$, as promised. 

Qualitatively, the stiffness and Coulomb energies increase with $\xi$, whereas the anisotropic contribution decreases with $\xi$. Thus, the optimal configuration with the lowest energy will have a nonzero $\xi$, as we will see in the next section. 

\section{Skyrmion Phases}\label{sec:5}

\begin{figure*}
    \centering \includegraphics[width=0.96\textwidth,height=0.47\textwidth]{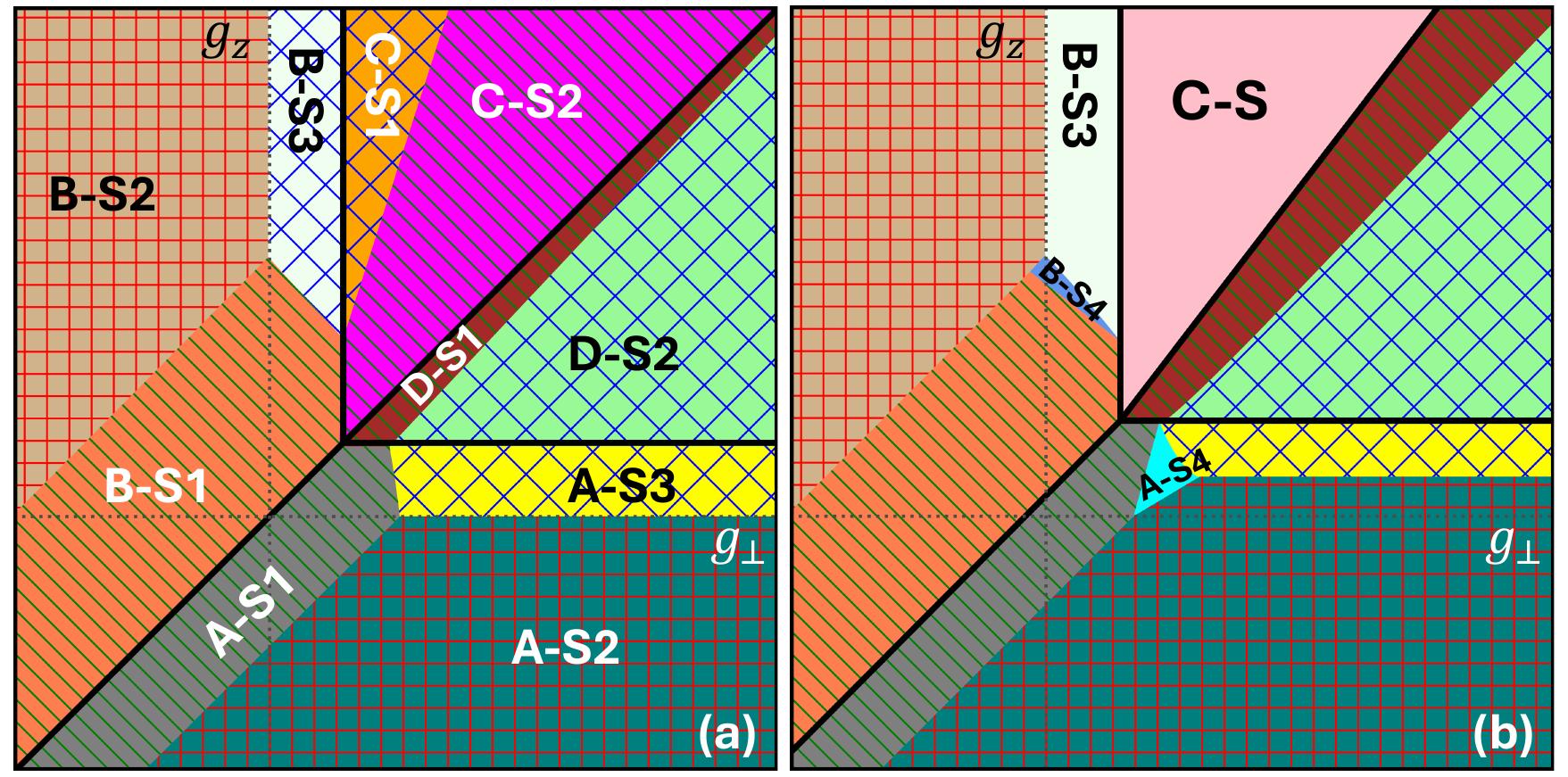}
\captionsetup{width=0.97\textwidth,
        justification=justified,
        singlelinecheck=false }
    \caption{Skyrmion phases above the  $\nu=-1$ QHFM ground states with $ g_{\perp}\in [-400,800]\text{meV}\cdot \text{nm}^{2}$, $ g_{z}\in [-400,800]\text{meV}\cdot \text{nm}^{2}$. The one-body fields are fixed at  $E_Z=1e_Z\ \&\ E_V=0$ in panel (a), and at  $E_Z=1e_Z\ \&\ E_V=0.3e_Z$ in panel (b). At $E_V=0$, there are ten skyrmion phases: A-S1-S3, B-S1-S3, C-S1-S2, \& D-S1-S2. At finite $E_V$, two new skyrmion phases, A-S4 and B-S4, appear, while C-S1 and C-S2  merge into a new skyrmion phase C-S. The solid black lines indicate the boundaries between the QHFM ground state phases. Phases with $\Xi=2$ in Eq.~\eqref{eq:relative_orientation} are hatched with red `+' symbols, phases with $\Xi=0$ are hatched with blue `x' symbols, and phases with $\Xi=-2$ are hatched with green `\textbackslash\textbackslash' symbols. The unhatched phases at $E_V\neq 0$ do not have quantized $\Xi$. In such skyrmion phases, the value of $\Xi$ varies continuously as couplings are varied. }
    \label{fig_skyrmion_phase_diagrams}
\end{figure*}

By minimizing the skyrmion energy functional of Eq.~\eqref{eq:skyrmion_varitational_energy} over the parameter space $(g_\perp,g_z)$, we get all possible skyrmion excited states above the various $\nu=-1$ QHFM ground states. It should be noted that a particular QHFM ground state can have multiple types of skyrmions, depending on the $SU(4)$ orientation of the central spinor $|f_2\rgl$. We choose the lowest energy skyrmion out of these possibilities to build the skyrmion phase diagrams shown in Fig.~\ref{fig_skyrmion_phase_diagrams}. 
To distinguish the various skyrmion phases quantitatively with a simple measure, we use the discriminator proposed previously~\cite{Lian_Rosch_Goerbig_2016, Lian_Goerbig_2017}. $\Xi$ is constructed from the relative orientations of the spin and valley order parameters at the center of the skyrmion and at infinity. It is defined as follows
\bean\label{eq:relative_orientation}
\hspace{-0.8cm}&&\ \Xi=\dfrac{\vec\tau_0\cdot \vec\tau_\infty}{|\vec\tau_0||\vec\tau_\infty|}-\dfrac{\vec\sigma_0\cdot \vec\sigma_\infty}{|\vec\sigma_0||\vec\sigma_\infty|},\nn
\hspace{-0.8cm}&&\vec\tau_\infty=\lgl\chi(\infty)|\vec\tau|\chi(\infty)\rgl=\big(\tau^x_{11},\tau^y_{11},\tau^z_{11}\big),\nn
\hspace{-0.8cm}&&\vec\tau_0=\lgl\chi(0)|\vec\tau|\chi(0)\rgl=\big(\tau^x_{22},\tau^y_{22},\tau^z_{22}\big),\nn
\hspace{-0.8cm}&&\vec\sigma_\infty=\lgl\chi(\infty)|\vec\sigma|\chi(\infty)\rgl=\big(\sigma^x_{11},\sigma^y_{11},\sigma^z_{11}\big),\nn
\hspace{-0.8cm}&&\vec\sigma_0=\lgl\chi(0)|\vec\sigma|\chi(0)\rgl=\big(\sigma^x_{22},\sigma^y_{22},\sigma^z_{22}\big)
\eean 
As can be seen later, in the case that both $|f_1\rgl\ \&\ |f_2\rgl$ are direct product states of valley and spin Bloch vectors, $\Xi=-2$ corresponds to a complete valley flip, $\Xi=2$ corresponds to a complete spin flip, while $\Xi=0$ corresponds to completely flipping both valley and spin.

\subsection{Skyrmions in Phase A}
We first consider the phase-A QHFM background of Eq.~\eqref{eq:phase_A_spinor}, with the spinor $|K,\upa\rgl$ occupied,  which implies that the spin and valley polarizations at infinity are
\bean
\vec\tau^A_{\infty}=\vec\sigma^A_{\infty}=\big(0,\ 0,\ 1\big).
\eean
We find four different skyrmions in phase A, labeled by A-S1, A-S2, A-S3, \& A-S4, each with a different central spinor.  As exhibited in Fig.~\ref{fig_ground_state_phase_diagram}(a)\&(b), skyrmions A-S1, A-S2, and A-S3 are the lowest energy skyrmions in certain regions of the couplings even when $E_V=0$. However, it is only when $E_V>0$ that the A-S4 skyrmion becomes the lowest energy configuration in a tiny region of couplings. Here are the central spinors corresponding to each type of skyrmion for the QHFM ground state A.
\begin{itemize}[left=0pt]
\item A-S1 has the central spinor
\bean
|f_2^\text{A-S1}\rgl=|K^\prime,\upa\rgl.
\eean 
This indicates a complete valley flip (as labeled by $\Xi^\text{A-S1}=-2$ in Fig.~\ref{fig_ground_state_phase_diagram}). 

\item A-S2 has the central spinor
\bean
|f_2^\text{A-S2}\rgl=|K,\dwa\rgl.
\eean
This indicates a complete spin flip (as labeled by $\Xi^\text{A-S2}=2$ in Fig.~\ref{fig_ground_state_phase_diagram}). 
\item A-S3 has the central spinor
\bean
|f_2^\text{A-S3}\rgl=|K^\prime,\dwa\rgl.
\eean 
This indicates a complete flip in both valley and spin (as labeled by $\Xi^\text{A-S3}=0$ in Fig.~\ref{fig_ground_state_phase_diagram}). 
\item A-S4 only appears at $E_V\neq0$ and has the central  spinor
\bean
|f_2^\text{A-S4}\rgl=\cos\dfrac{\alpha}{2}|K,\dwa\rgl+\sin\dfrac{\alpha}{2}|K^\prime,\upa\rgl,
\eean 
where the nontrivial angle $\alpha$ can only be obtained numerically.  The resulting valley and spin polarizations are 
\bean
\vec\tau_0^\text{A-S4}=-\vec\sigma_0^\text{A-S4}=\big(0,\ 0,\ \cos\alpha\big),
\eean 
leading to $\Xi^\text{A-S4}=2\cos\alpha/|\cos\alpha|=\pm 2$ depending on the sign of $\cos\alpha$. The skyrmion phase A-S4 appears to have been missed in previous work~\cite{Lian_Rosch_Goerbig_2016,Lian_Goerbig_2017}.
\end{itemize}
\subsection{Skyrmions in Phase B}
Now we consider the QHFM background to be phase B, Eq.~\eqref{eq:phase_B_spinor}, with the occupied spinor $|\btau,\upa\rgl$. This leads to the following spin and valley polarizations at infinity
\bean
\vec\tau^B_{\infty}=\big(\sin\theta_\tau,\ 0,\ \cos\theta_\tau\big),\quad\vec\sigma^B_{\infty}=\big(0,\ 0,\ 1\big),
\eean
we find four different skyrmions labeled by B-S1, B-S2, B-S3 and B-S4. As seen in Fig.~\ref{fig_ground_state_phase_diagram}(a)\&(b), the first three skyrmion phases are seen even for $E_V=0$, whereas the skyrmion phase B-S4 is only stabilized for $E_V>0$. Here are the central spinors of the various skyrmion phases:
\begin{itemize}[left=0pt]
\item B-S1  has the central spinor
\bean
|f_2^\text{B-S1}\rgl=|-\btau,\upa\rgl.
\eean 
This represents a complete valley flip (labeled by $\Xi^\text{B-S1}=-2$ in Fig.~\ref{fig_ground_state_phase_diagram}). 
\item B-S2  has the central spinor
\bean
|f_2^\text{B-S2}\rgl=|\btau,\dwa\rgl,
\eean 
which indicates a complete valley flip (as labeled by $\Xi^\text{B-S2}=2$ in Fig.~\ref{fig_ground_state_phase_diagram}). 
\item B-S3 has the central spinor
\bean
|f_2^\text{B-S3}\rgl=|-\btau^\prime,\dwa\rgl,\ \lvert-\btau^\prime\rangle=\begin{pmatrix}
    \sin\frac{\theta_{\tau^\prime}}{2}\\
    -\cos\frac{\theta_{\tau^\prime}}{2}
\end{pmatrix},
\eean 
where $\theta_{\tau^\prime}$ can only be determined numerically. We find that $\theta_{\tau^\prime}=\theta_\tau=\frac{\pi}{2}$ at $E_V=0$. The resulting valley and spin polarizations are 
\bean
&&\vec\tau_0^\text{B-S3}=-\big(\sin\theta_{\tau^\prime},\ 0,\ \cos\theta_{\tau^\prime}\big),\nn
&&\vec \sigma_0^\text{B-S3}=\big(0,\ 0,\ -1\big),
\eean 
leading to $\Xi^\text{B-S3}=1-\cos(\theta_\tau-\theta_{\tau^\prime})$. When $E_V=0$, we find $\Xi=0$ as shown in Fig.~\ref{fig_skyrmion_phase_diagrams}(a), representing a complete flip in both valley and spin.
\item B-S4  only appears at $E_V\neq0$ and has the central spinor
\bean
|f_2^\text{B-S4}\rgl=\cos\dfrac{\alpha}{2}|-\btau,\upa\rgl+\sin\dfrac{\alpha}{2}|\btau^\prime,\dwa\rgl,
\eean 
where $\alpha$ and $\theta_{\tau^\prime}$ can only be determined numerically.  The resulting valley and spin polarizations are 
\bean
&&\vec\tau_0^\text{B-S4}=\big(\sin^2\dfrac{\alpha}{2}\sin\theta_{\tau^\prime}-\cos^2\dfrac{\alpha}{2}\sin\theta_\tau,\ 0,\nn 
&&\quad\quad\quad\quad\ \sin^2\dfrac{\alpha}{2}\cos\theta_{\tau^\prime}-\cos^2\dfrac{\alpha}{2}\cos\theta_\tau\big),\nn
&&\vec\sigma_0^\text{B-S4}=\big(\sin\alpha\sin(\theta_\tau-\theta_{\tau^\prime}),\ 0,\ \cos\alpha\big).
\eean 
We will not present the expression for $\Xi$, which is not illuminating, but simply note that it is not quantized. The skyrmion phase B-S4 appears to have been missed in previous work~\cite{Lian_Rosch_Goerbig_2016,Lian_Goerbig_2017}.
\end{itemize}
\subsection{Skyrmions in Phase C}

Next, we consider skyrmion excitations above phase C of the QHFM, with the occupied spinor $\cos\dfrac{\alpha}{2}|\btau,\upa\rgl-\sin\dfrac{\alpha}{2}|-\btau,\dwa\rgl$ [Eq.~\eqref{eq:phase_C_spinor}]. The  spin and valley polarizations at  infinity are
\bean
\hspace{-0.2cm}\vec\tau^C_{\infty}=\cos\alpha\big(\sin\theta_\tau,\ 0,\ \cos\theta_\tau\big),\ \vec\sigma^C_{\infty}=\big(0,\ 0,\ \cos\alpha\big),
\eean
where $\alpha$ and $\theta_\tau$ in $\btau$ are given in Eq.~\eqref{eq:phase_C_spinor}.
 
The lowest energy skyrmion has the central spinor
\bean\label{eq:C-S_spinor}
\hspace{-0.5cm}&&|f_2^\text{C-S}\rgl=\cos\dfrac{\theta}{2}\Big(\sin\dfrac{\alpha}{2}|\btau,\upa\rgl+\cos\dfrac{\alpha}{2}|-\btau,\dwa\rgl\Big)\nn
\hspace{-0.5cm}&&\quad\quad\quad+\sin\dfrac{\theta}{2}\Big(\sin\dfrac{\alpha^\prime}{2}|-\btau,\upa\rgl+\cos\dfrac{\alpha^\prime}{2}|\btau,\dwa\rgl\Big),
\eean 
where $\theta,\ \alpha^\prime$ can only be determined numerically. This is generically true when $E_V>0$. However,  in the limit of $E_V\rightarrow0$, the angles acquire special values, and the central spinors become simpler.  This leads to the skyrmion phases C-S1 and C-S2 shown in Fig.~\ref{fig_ground_state_phase_diagram}(a) and (b).
\begin{itemize}[left=0pt]
\item C-S1 only occurs at $E_V=0$, with the angles taking the values $\theta=0,\ \btau=\hat e_x$, leading to the central spinor
\bean
|f_2^\text{C-S1}\rgl=\sin\dfrac{\alpha}{2}|\hat e_x,\upa\rgl+\cos\dfrac{\alpha}{2}|-\hat e_x,\dwa\rgl.
\eean 
This implies the valley and spin polarizations at the center
\bean
\vec\tau_0^\text{C-S1}=\big(-\cos\alpha,\ 0,\ 0\big),\ 
\vec\sigma_0^\text{C-S1}=\big(0,\ 0,\ -\cos\alpha\big),
\eean 
thus the resulting $\Xi^\text{C-S1}=0$.
\item C-S2 also only occurs at $E_V=0$, with the angles taking the special values $\theta=\pi,\ \btau=\hat e_x$, leading to the central spinor
\bean
|f_2^\text{C-S2}\rgl=\sin\dfrac{\alpha^\prime}{2}|-\hat e_x,\upa\rgl+\cos\dfrac{\alpha^\prime}{2}|\hat e_x,\dwa\rgl. 
\eean 
This implies the valley and spin polarizations at the center
\bean
\vec\tau_0^\text{C-S1}=\big(\cos\alpha^\prime,\ 0,\ 0\big),\ 
\vec\sigma_0^\text{C-S1}=\big(0,\ 0,\ -\cos\alpha^\prime\big),
\eean 
resulting in $\Xi^\text{C-S2}=-2$.
\end{itemize}
It should be noted that there are other skyrmions in phase C. However, since they never have the lowest energy for any set of couplings, we ignore them. 

\subsection{Skyrmions in Phase D}
Finally, we consider skyrmion excitations above phase D of the QHFM, with the occupied spinor \newline $\cos\dfrac{\alpha}{2}|K,\upa\rgl+\sin\dfrac{\alpha}{2}|K^\prime,\dwa\rgl$, where the angle $\alpha$ is defined in Eq.~\eqref{eq:phase_D_spinor}.  The valley and spin polarizations at infinity are
\bean
\vec\tau^D_{\infty}=\vec\sigma^D_{\infty}=\big(0,\ 0,\ \cos\alpha\big).
\eean
We find two lowest energy skyrmions labeled D-S1 and  D-S2 as shown in Fig.~\ref{fig_ground_state_phase_diagram}(a)\&(b).
\begin{itemize}[left=0pt]
\item D-S1 has the central spinor 
\bean
|f_2^\text{D-S1}\rgl=\cos\dfrac{\alpha^\prime}{2}|K,\dwa\rgl+\sin\dfrac{\alpha^\prime}{2}|K^\prime,\upa\rgl.
\eean
This leads to central spin and valley polarizations
\bean
\vec \tau^\text{D-S1}_0=-\vec\sigma_0^\text{D-S1}=\big(0,\ 0,\ \cos\alpha^\prime\big),
\eean 
resulting  in $\Xi^\text{D-S1}=-2$.
\item D-S2 has the central spinor 
\bean
|f_2^\text{D-S2}\rgl=\sin\dfrac{\alpha}{2}|K,\upa\rgl-\cos\dfrac{\alpha}{2}|K^\prime,\dwa\rgl. 
\eean
This leads to spin and valley polarizations
\bean
\vec \tau^\text{D-S2}_0=\vec\sigma_0^\text{D-S2}=\big(0,\ 0,\ -\cos\alpha\big),
\eean 
resulting in  $\Xi^\text{D-S2}=0$.
\end{itemize}

\subsection{Energy and Size of the Skyrmions}

\begin{figure*}
    \centering \includegraphics[width=0.97\textwidth,height=0.45\textwidth]{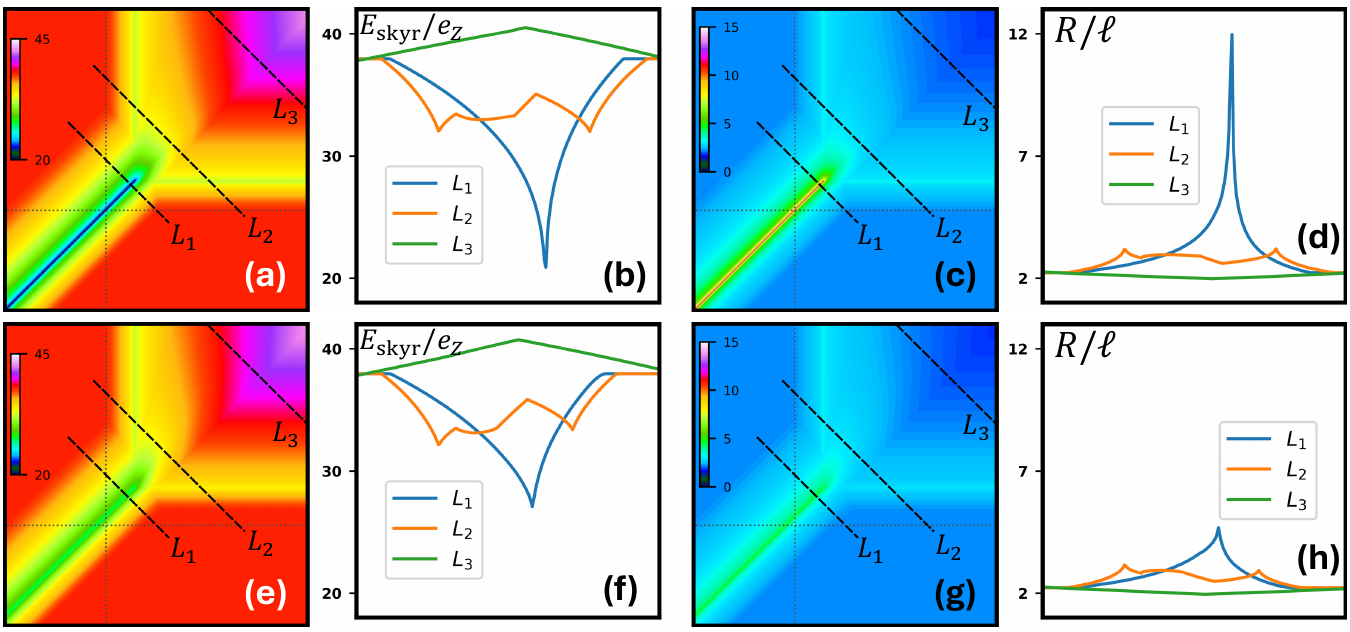}
    \captionsetup{width=0.97\textwidth,
        justification=justified,
        singlelinecheck=false }
    \caption{(a) Skyrmion energy $E_\text{skyr}$ (in unit of $e_Z$) over a range of couplings for $E_V=0$, corresponding to the phase diagram of Fig.~\ref{fig_skyrmion_phase_diagrams}{(a)}. (b) The skyrmion energy of panel (a) is plotted along the three different sections $L_1,\ L_2\ \&\ L_3$, shown in panel (a). (c)  The size $R$ of the optimal skyrmion (in units of $\ell$) over a range of couplings for $E_V=0$, also associated with the skyrmion phase diagram Fig.~\ref{fig_skyrmion_phase_diagrams}{(a)}. (d) The size of the skyrmion plotted along the three sections  $L_1,\ L_2\ \&\ L_3$ of panel (a). (e) Skyrmion energy for a range of couplings for $E_V=0.3E_Z$, corresponding to the phase diagram of Fig.~\ref{fig_skyrmion_phase_diagrams}{(b)}. (f) The skyrmion energy in units of $e_Z$ plotted along the sections $L_1,\ L_2\ \&\ L_3$ of panel (e).  (g) The size of the optimal skyrmion for $E_V=0.3E_Z$, corresponding to the phase diagram of Fig.~\ref{fig_skyrmion_phase_diagrams}{(b)}. The size of the optimal skyrmion is plotted along the three sections $L_1,\ L_2\ \&\ L_3$ of panel (e). The values of $(g_\perp,g_z)$ in unit of $\text{meV}\cdot \text{nm}^{2}$ for three lines are $L_1:(-150,350)\rightarrow(250,-50)$, $L_2:(-50,575)\rightarrow(550,-25)$, $L_3:(400,800)\rightarrow(800,400)$. }
    \label{fig_skyrmion_energy_size}
\end{figure*}

In panels (a) and (e) of Fig.~\ref{fig_skyrmion_energy_size}, we present heat maps of the skyrmion energy $E_\text{skyr}$, corresponding to the skyrmion phase diagrams shown in Fig.~\ref{fig_skyrmion_phase_diagrams}(a) and (b). In Fig.~\ref{fig_skyrmion_energy_size}(b) and (f), we plot the skyrmion energy along three representative lines labeled by $L_1,\ L_2\ \&\ L_3$, indicated in panels (a) and (e).

The skyrmion size $R$ is defined as the average of $r$ over the topological density $\varrho(\br)$~\cite{Lian_Goerbig_2017}, 
\bean\label{eq:skyrmion_size}
R=\int r\varrho(\br)d^2\br=\int drrW^\prime(r)=\dfrac{\sqrt{\pi}\Gamma(\xi+1/2)}{2\Gamma(\xi+1)}\lambda.
\eean
In Figs.~\ref{fig_skyrmion_energy_size}(c) and (g), we present heat maps of the optimal skyrmion size $R$, corresponding to the skyrmion phase diagrams shown in Fig.~\ref{fig_skyrmion_phase_diagrams}. In Fig.~\ref{fig_skyrmion_energy_size}(d) and (h), we plot the skyrmion size along three representative lines labeled by $L_1,\ L_2\ \&\ L_3$, indicated in panels (c) and (g).\\
From Figs.~\ref{fig_skyrmion_energy_size}(a), (c), (e), and (g), we observe that the skyrmion energy reaches its minimum, while the skyrmion size attains its maximum, along the phase boundary separating the A-S1 and B-S1 skyrmion phases in Fig.~\ref{fig_skyrmion_phase_diagrams}. This boundary corresponds to the phase boundary between phases A and B in the QHFM ground-state phase diagram shown in Fig.~\ref{fig_ground_state_phase_diagram}. Equating the ground-state energies $E^\text{A}_\text{g.s.}$ and $E^\text{A}_\text{g.s.}$ given in Eqs.~\eqref{eq:phase_A_spinor} and \eqref{eq:phase_B_spinor}, we obtain the phase boundary
\bean
(u_{z,H}-u_{z,F})=E_V+(u_{\perp,H}-u_{\perp,F}),
\eean 
which, upon using Eq.~\eqref{eq:u1_HF_couplings}, can be rewritten as $g_z=\pi\ell^2E_V+g_\perp$. As we approach this boundary,
\bean
|f_1^\text{A}\rgl=|f_1^\text{B}\rgl=|K,\upa\rgl,\quad |f_2^\text{A-S1}\rgl=|f_2^\text{B-S1}\rgl=|K^\prime,\upa\rgl,
\eean 
with which, $\cE_1\ \&\ \cE_2$ defined in Eqs.~\eqref{eq:skyr_E1}, \eqref{eq:skyr_E2} reduce to $\cE_1=2E_V,\ \cE_2=0$. The skyrmion energy, therefore, simplifies to
\bean
E_\text{skyr}=\pi\rho_sF(\xi)+\dfrac{e^2X(\xi)}{2\epsilon\lambda}+\dfrac{\lambda^2}{\ell^2}\dfrac{E_V}{\xi},
\eean
whose variational minimization shows that, in the limit  $E_V\rightarrow 0$, the optimal parameters approach $\xi=0,\ \lambda\rightarrow\infty$, which lead to a skyrmion with lowest energy $4\pi\rho_s$ and divergent size. (In the practical numerical calculations, the limit $E_V\rightarrow0$ is implemented by introducing a small but finite value, $E_V\sim 10^{-6}$.)

\section{Robustness with respect to the function $W(r)$}\label{sec:6}
The specific form of the radial profile $W(r)$ is not unique. Indeed, the variational ansatz introduced in the main text represents a particular one-parameter family of skyrmion textures that satisfy the boundary conditions in Eq.~\eqref{eq:W_conditions}. More generally, any radial profile that interpolates smoothly between $W(0)=1$ and $W(\infty)=0$ is physically admissible.  The detailed functional form of $W(r)$ is expected to be determined by the competition among the various energy contributions. Since one needs an infinite number of parameters to specify a function, it is not possible to do the calculation for a completely general $W(r)$.  However, we can formulate the variational problem in a form that is independent of any particular choice of ansatz. Since $W(r)$ is dimensionless, whereas $r$ carries the dimension of length, dimensional analysis implies that the most general implicit form of $W(r)$ can always be expressed as  
\bean
W(r)=W(r;\lambda,\vec\xi),\nonumber
\eean 
where $\lambda$ is the characteristic length needed to balance the dimension of $r$, while $\vec\xi=(\xi_1,\xi_2,...)$ denotes an, in general, infinite set of dimensionless variational parameters characterizing the function $W(r)$. Regardless of the specific functional form of $W(r)$, dimensional analysis determines the scaling of each contribution to the skyrmion energy functional,
\bean
\hspace{-1cm}&&\text{Stiffness and Coulomb energies:}\nn
\hspace{-1cm}&&\quad \quad E_\text{stiff}\sim \rho_s F(\vec\xi),\quad E_\text{Coul}\sim\dfrac{e^2}{\varepsilon\lambda}X(\vec \xi),\nn
\hspace{-1cm}&&\text{Symmetry breaking energy:}\nn
\hspace{-1cm}&&\quad \quad E_\text{SB}\sim \dfrac{\lambda^2}{\ell^2}\bigg[\cE_1 Y_1(\vec\xi)+\cE_2 Y_2(\vec\xi)\bigg].
\eean 
where all functions about $\vec\xi$ are dimensionless and are given by
\bean\label{eq:integrals_general_W}
&&\pi F(\vec \xi)=\int d^2\br\bigg[\dfrac{2W\big(1-W\big)}{r^2}+\dfrac{W^{\prime2}}{2W\big(1-W\big)}\bigg],\nn
&&\dfrac{1}{\lambda}X(\vec\xi)=\int d^2\br_1d^2\br_2\dfrac{W^\prime(r_1)}{2\pi r_1}\dfrac{1}{|\br_1-\br_2|}\dfrac{W^\prime(r_2)}{2\pi r_2},\nn
&&\lambda^2Y_1(\vec\xi)=\int d^2\br W,\quad \lambda^2Y_2(\vec\xi)=\int d^2\br W^2.
\eean 
We find that the optimal skyrmion is relatively insensitive to the choice of different functional forms for the radial ansatz $W(r)$. To illustrate this, we compare the results obtained using Eq.~\eqref{eq:w(r)} with integer values of $\xi$, as well as those obtained from other functional forms of $W(r)$ not contained in the one-parameter family $W_\xi(r)$.

\subsection{$W_\xi(r)$ with integer values of $\xi$}
\begin{figure}[h]
    \centering \includegraphics[width=0.48\textwidth,height=0.45\textwidth]{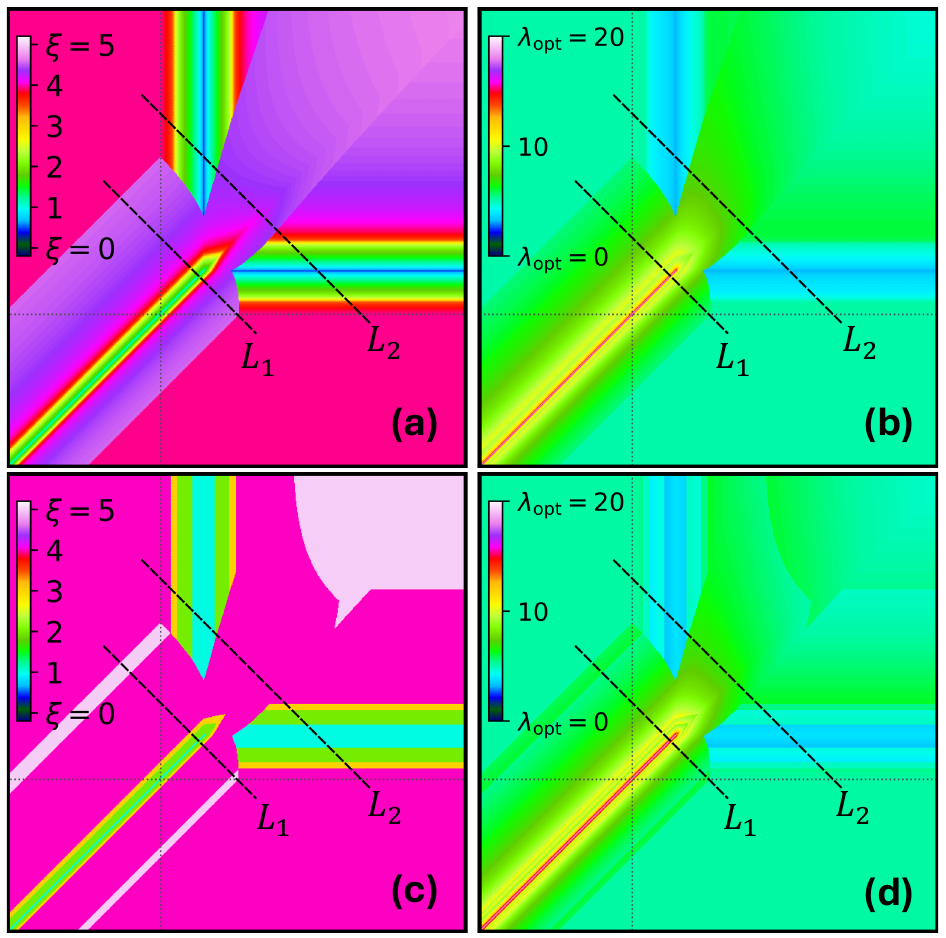}
    \caption{Comparison between the optimal continuous and integer values of $\xi$ for the ansatz in Eq.~\eqref{eq:w(r)}. (a) Heat map of the optimal $\xi$ taking continuous values, corresponding to skyrmion phase diagram Fig.~\ref{fig_skyrmion_phase_diagrams}\textcolor{red}{(a)}. (b) Optimal characteristic length parameter $\lambda_{\mathrm{opt}}$ associated with the continuous optimal $\xi$. (c) Heat map of the optimal $\xi$ when $\xi$ is restricted to integer values. (d) Optimal characteristic length $\lambda_{\mathrm{opt}}$ associated with the optimal integer $\xi$. The lines $L_1$ and $L_2$ are identical to those shown in Fig.~\ref{fig_skyrmion_energy_size} and will be used in Fig.~\ref{fig_L1_int}.
    }
    \label{fig_xi_lmd}
\end{figure}

\begin{figure}[h]
    \centering \includegraphics[width=0.46\textwidth,height=0.49\textwidth]{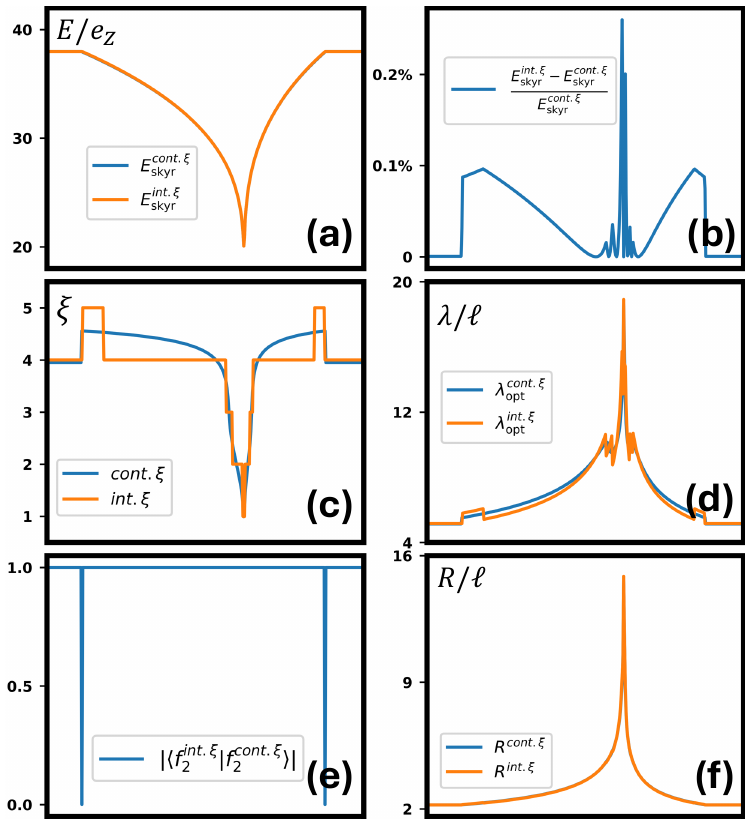}
    \caption{Comparison between the optimal skyrmions resulting from Eq.~\eqref{eq:w(r)} with continuous and integer $\xi$ along line $L_1$ in Fig.~\ref{fig_skyrmion_energy_size}. (a) Skyrmion energies $E^{\text{cont.}~\xi}_\text{skyr}$ for continuous $\xi$ and $E^{\text{int.}~\xi}_\text{skyr}$ for integer $\xi$ along $L_1$. The two lines are indistinguishable to the eye. (b) Relative difference between $E^{\text{cont.}~\xi}_\text{skyr}$ and $E^{\text{int.}~\xi}_\text{skyr}$ plotted along $L_1$. The relative difference is always less than 0.5\%. (c) Optimal values of the continuous and integer $\xi$ along $L_1$. (d) Optimal characteristic length $\lambda_{\mathrm{opt}}$ associated with the continuous and integer $\xi$ along $L_1$. The two curves are extremely close to each other. (e) The overlap $\langle f^{\text{cont.}~\xi}_2|f^{\text{int.}~\xi}_2\rgl$ between the spinors in the skyrmion core.  $|f^{\text{cont.}~\xi}_2\rgl$ corresponds to continuous $\xi$ and $|f^{\text{int.}~\xi}_2\rgl$ corresponds to integer $\xi$. (f) Skyrmion size (defined by Eq.~\eqref{eq:skyrmion_size}) of the optimal skyrmions arising from continuous and integer $\xi$. The two curves are indistinguishable to the eye. 
    }
    \label{fig_L1_int}
\end{figure}

\begin{figure}[h]
    \centering \includegraphics[width=0.46\textwidth,height=0.49\textwidth]{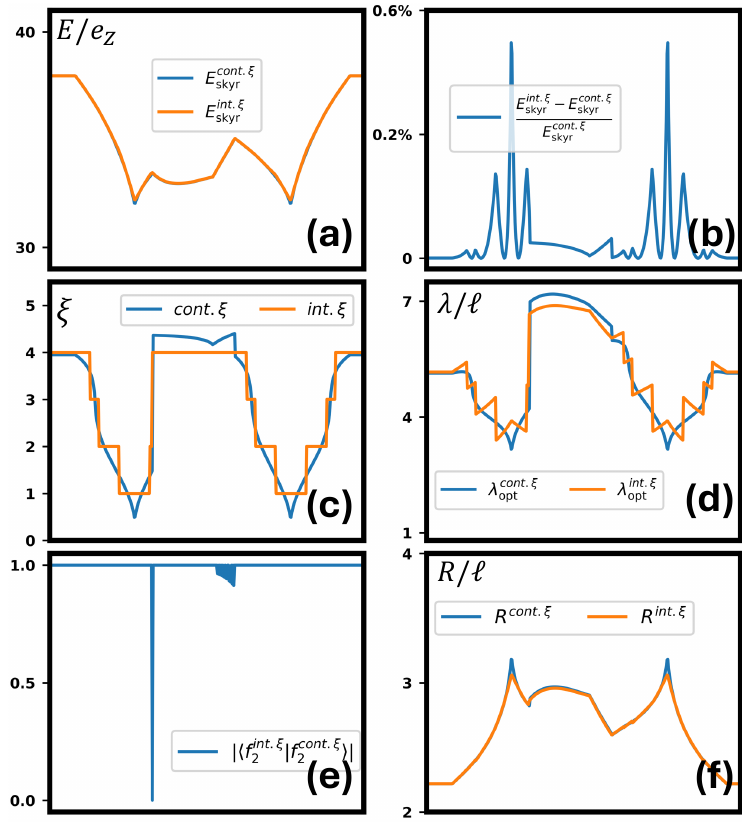}
    \caption{Comparison of optimal skyrmions resulting from Eq.~\eqref{eq:w(r)} with continuous and integer $\xi$ along the line $L_2$ in Fig.~\ref{fig_skyrmion_energy_size}. (a) Skyrmion energies $E^{\text{cont.}~\xi}_\text{skyr}$ for continuous $\xi$ and $E^{\text{int.}~\xi}_\text{skyr}$ for integer $\xi$. The two curves are nearly indistinguishable to the eye. (b) The relative difference between $E^{\text{cont.}~\xi}_\text{skyr}$ and $E^{\text{int.}~\xi}_\text{skyr}$ remains at most about 0.5\%. (c) Optimal values of the continuous and integer $\xi$. (d) Optimal characteristic length $\lambda_\text{opt}$ associated with the optimal continuous and integer $\xi$. (e) The overlap $\langle f^{\text{cont.}~\xi}_2|f^{\text{int.}~\xi}_2\rgl$ between the spinors in the skyrmion center.  $|f^{\text{cont.}~\xi}_2\rgl$, arises from continuous $\xi$ and $|f^{\text{int.}~\xi}_2\rgl$ arises from integer $\xi$. (f) Sizes of the optimal skyrmions resulted from continuous and integer $\xi$. The curves are nearly indistinguishable to the eye. 
    }
    \label{fig_L2_int}
\end{figure}

We first restrict the variational parameter $\xi$ in Eq.~\eqref{eq:w(r)} to positive integers. This has the advantage that the function $F(\xi)$, required for the stiffness energy, can be computed analytically, as shown in Eq.~\eqref{eq:F for integer xi}. For each value of the couplings $g_z,\ g_\perp$, we allow $\xi$ to take integer values and choose the integer which leads to the lowest total energy. The energies of the skyrmions with optimal continuous and integer values of $\xi$,  together with their corresponding optimal characteristic lengths $\lambda_\text{opt}$, are presented in Fig.~\ref{fig_xi_lmd}. The optimal continuous $\xi$ and the associated $\lambda_\text{opt}$, corresponding to the phase diagram in Fig.~\ref{fig_skyrmion_phase_diagrams}{(a)}, are shown in Fig.~\ref{fig_xi_lmd}(a) and (b), respectively. The corresponding results obtained with integer-valued $\xi$ are shown in Fig.~\ref{fig_xi_lmd}(c) and (d), respectively.\\
To demonstrate the agreement between the two variational approaches quantitatively, we make a detailed comparison along the lines $L_1$ and  $L_2$ highlighted in Fig.~\ref{fig_xi_lmd}, exhibited in Figs.~\ref{fig_L1_int} and \ref{fig_L2_int}. In panels (a) of Figs.~\ref{fig_L1_int} and \ref{fig_L2_int}, we plot the skyrmion energies obtained from the ansatz in Eq.~\eqref{eq:w(r)} using continuous and integer-valued $\xi$ along lines $L_1$ and $L_2$, respectively. The difference between the optimal energies obtained using continuous values of $\xi$ and integer values of $\xi$ is indistinguishable to the eye. In Figs.~\ref{fig_L1_int}(b) and \ref{fig_L2_int}(b) we plot the relative difference of energies between the integer and continuous $\xi$ ans\"atze to find that it never exceeds 1\%. Next, in Figs.~\ref{fig_L1_int}(c) and \ref{fig_L2_int}(c) we plot the optimal values of $\xi$ found in the two ans\"atze, while Figs.~\ref{fig_L1_int}(d) and \ref{fig_L2_int}(d) show the corresponding characteristic lengths $\lambda$. In panels Figs.~\ref{fig_L1_int}(e) and \ref{fig_L2_int}(e), we show the overlap of the central spinor of the optimal skyrmion for the integer and continuous $\xi$ ans\"atze, which are essentially unity, except near the phase transition between different skyrmion types. Finally, in Figs.~\ref{fig_L1_int}(f) and \ref{fig_L2_int}(f) we show the average size of the optimal skyrmions in the two ans\"atze, with the difference again being nearly invisible to the eye. Overall, we find that the relative difference of energies never exceeds 1\% anywhere in the $(g_\perp,\ g_z)$ parameter space. \\

\subsection{Ans\"atze with functional forms different from $W_\xi(r)$}

\begin{figure}[h]
    \centering \includegraphics[width=0.46\textwidth,height=0.35\textwidth]{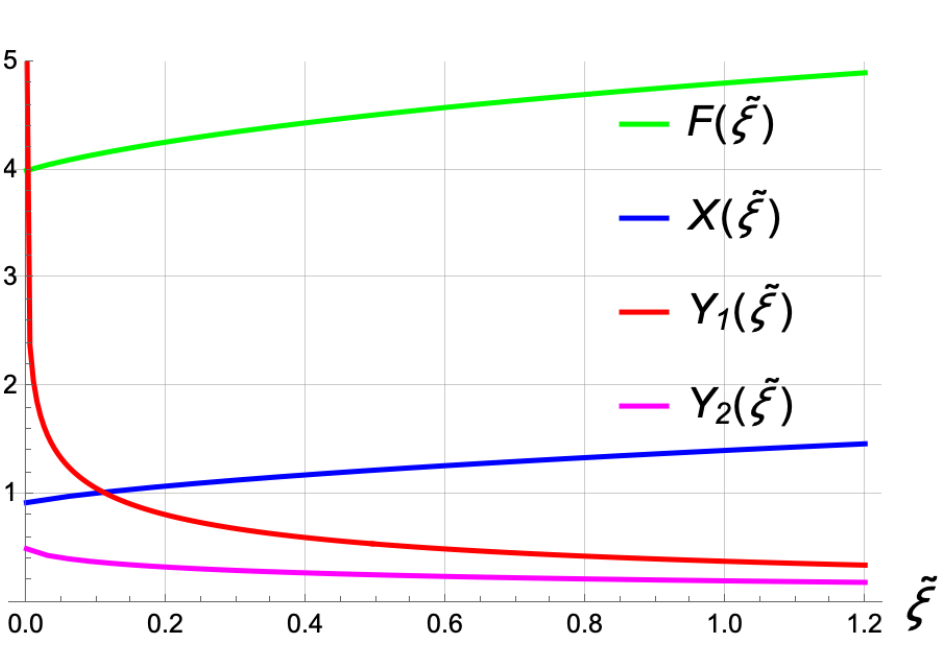}
    \caption{Dependence of functionals $F,\ X,\ Y_1\ \&\ Y_2$ in Eq.~\eqref{eq:app_varitational_energy_tilde_W} on variational parameter $\tilde\xi$. $F\ \&\ X$ increase with $\tilde\xi$ (similar to Fig.~\ref{fig_ft_X_xi}) while $Y_1\ \&\ Y_2$ decrease with $\tilde\xi$ (similar to Eqs.~\eqref{eq:app_integral_over_W}\&\eqref{eq:app_integral_over_W_sqrt}).
    }
    \label{fig_integrals_tilde_W}
\end{figure}

In Refs.~\cite{Lian_Rosch_Goerbig_2016, Lian_Goerbig_2017}, a deformation of the BPS profile is introduced by promoting the length parameter $\lambda$ in $W_0$, Eq.~\eqref{eq:w_bps}, to an $r$-dependent function, 
\bean
\lambda\ \rightarrow\ \lambda(r)=\lambda_0 \exp\left(-\dfrac{r^2}{\kappa\lambda^2_0}\right),
\eean 
where $\kappa$ is a variational parameter. Defining
 $\tilde\xi={2}/{\kappa}$, and omitting the subscript on $\lambda_0$ for notational simplicity, the corresponding radial profile can be written as
\bean\label{eq:app_ansatz_tilde_W}
\tilde W(r; \tilde\xi)=\dfrac{\lambda^2}{\lambda^2+e^{\tilde\xi r^2/\lambda^2}r^2},
\eean 
with which the skyrmion energy functional reads
\bean\label{eq:app_varitational_energy_tilde_W}
&&E^{\tilde W}_\text{skyr}=\pi\rho_sF(\tilde\xi)+\dfrac{e^2}{2\epsilon\lambda}X(\tilde\xi)\nn
&&\quad\quad\quad +\dfrac{\lambda^2}{\ell^2}\cE_1Y_1(\tilde\xi)+\dfrac{\lambda^2}{\ell^2}\cE_2Y_2(\tilde\xi),
\eean 
where $F,\ X,\ Y_1,\ Y_2$ are dimensionless functionals of $\tilde W$ defined in Eq.~\eqref{eq:integrals_general_W}. These functionals can only be evaluated numerically for this functional form of $W(r)$, with the resulting dependence on $\tilde\xi$ presented in Fig.~\ref{fig_integrals_tilde_W}. 

As in the case of our ansatz $W_\xi(r)$,  $F$ and $X$ increase monotonically with $\tilde\xi$, implying that the stiffness and Coulomb contributions to the skyrmion energy increase with $\tilde\xi$. In contrast, $Y_1$ and $Y_2$, which enter the symmetry-breaking energy terms, decrease monotonically with $\tilde\xi$, consistent with Eqs.~\eqref{eq:app_integral_over_W} and \eqref{eq:app_integral_over_W_sqrt}.  The competition between these two trends gives rise to an optimal value of $\tilde\xi$ that minimizes the total skyrmion energy.

\begin{figure}[h]
    \centering \includegraphics[width=0.46\textwidth,height=0.35\textwidth]{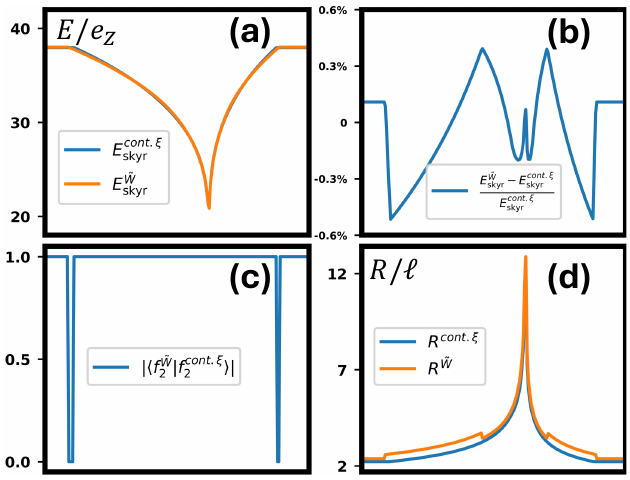}
    \caption{ Comparison between the  optimal skyrmions resulting from ansatz Eq.~\eqref{eq:w(r)} with continuous $\xi$ and ansatz Eq.~\eqref{eq:app_ansatz_tilde_W} along line $L_1$ in Fig.~\ref{fig_skyrmion_energy_size}. (a) Skyrmion energies of the two ans\"atze. The curves are nearly indistinguishable to the eye.  (b) The relative difference between $E^{\text{cont.}~\xi}_\text{skyr}$ and $E^{\tilde W}_\text{skyr}$ remains within about 0.5\%. (c) The overlap between the spinors of the two ans\"atze in the skyrmion core. (d) Sizes of the optimal skyrmions, $R^{\tilde W}$, in the two ans\"atze. The curves are quite close. 
    }
    \label{fig_L1_tilde_W}
\end{figure}
\begin{figure}[h]
    \centering \includegraphics[width=0.46\textwidth,height=0.35\textwidth]{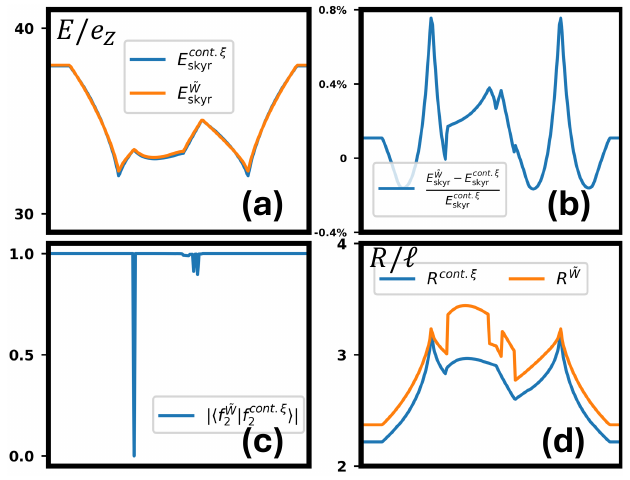}
    \caption{Comparison between the  optimal skyrmions resulting from ansatz Eq.~\eqref{eq:w(r)} with continuous $\xi$ and ansatz Eq.~\eqref{eq:app_ansatz_tilde_W} along line $L_2$ in Fig.~\ref{fig_skyrmion_energy_size}. (a) Skyrmion energies for the two ans\"atze. The curves are barely distinguishable. (b) The relative difference between the energies of the two ans\"atze stays within 1\%.  (c) The overlap between the optimal spinors of the two ans\"atze in the skyrmion core. (d) Sizes of the optimal skyrmions in the two ans\"atze.
    }
    \label{fig_L2_tilde_W}
\end{figure}

Figs.~\ref{fig_L1_tilde_W} and \ref{fig_L2_tilde_W} compare the skyrmion solutions obtained  from the ansatz $W_\xi(r)$ in Eq.~\eqref{eq:w(r)} with continuous $\xi$ and the ansatz $\tilde W(r;\tilde\xi)$ in Eq.~\eqref{eq:app_ansatz_tilde_W}. Panels (a) of Figs.~\ref{fig_L1_tilde_W} and \ref{fig_L2_tilde_W} 
show the skyrmion energies obtained from these two different ansatz along lines $L_1$ and $L_2$, respectively, while the corresponding relative energy differences are shown in panels (b). The two sets of results are in excellent agreement, with the relative difference remaining below 1\%. Indeed, we have verified that the relative difference never exceeds 1\% anywhere in the $(g_\perp,\ g_z)$ parameter space. \\
In panels (c) of Figs.~\ref{fig_L1_tilde_W} and \ref{fig_L2_tilde_W}, we plot the overlap between the spinors $|f_2^{\text{cont.}~\xi}\rgl$ and $|f_2^{\tilde W}\rgl$, where the latter denotes the spinor occupied at the skyrmion center for the optimal solutions obtained from the ansatz in Eq.~\eqref{eq:app_ansatz_tilde_W}, along lines $L_1$ and $L_2$, respectively. We find that the overlap $|\lgl f_2^{\tilde W}|f_2^{\text{cont.}~\xi}\rgl|$ remains close to unity over almost the entire lengths of $L_1$ and $L_2$, confirming the excellent agreement between the two variational ans\"{a}tze. Noticeable deviations from unity occur only in the immediate vicinity of the phase boundaries, indicating that the two variational approaches predict essentially identical skyrmion phase diagrams, differing only by negligible displacements of the phase boundaries.\\
In panels (d) of Figs.~\ref{fig_L1_tilde_W} and \ref{fig_L2_tilde_W}, we compare the skyrmion size $R$ of skyrmions resulting from these two variational ans\"{a}tze along lines $L_1$ and $L_2$, respectively.
We have also examined several alternative variational ans\"{a}tze closely related to Eq.~\eqref{eq:app_ansatz_tilde_W},
\bean
\dfrac{\lambda^2}{e^{\tilde\xi r^2/\lambda^2}\lambda^2+r^2},\quad \dfrac{\lambda^2}{e^{\tilde\xi r^2/\lambda^2}\big(\lambda^2+r^2\big)},
\eean 
which yield results in equally good agreement with those presented in Figs.~\ref{fig_L1_tilde_W} and \ref{fig_L2_tilde_W}. We therefore conclude that the variationally optimized skyrmion is robust with respect to the choice of the radial ansatz.

\section{Conclusion and Outlook}
\label{sec:conclusion}
Quantum Hall ferromagnetism~\cite{Shivaji_Skyrmion, QHFM_Yang_etal_1994, QHFM_Moon_etal_1995} is a spontaneous symmetry-breaking phenomenon which occurs whenever there are internal degrees of freedom, and whenever a manifold of Landau levels is partially full. Initially discovered in 2D electron gases confined in semiconductor heterostructures, in graphene, it leads to a plethora of possible phases due to the approximate $SU(4)$ symmetry of each Landau level manifold. 

A characteristic of quantum Hall ferromagnets is that they can support extended spin/valley textured excitations called skyrmions. Due to the spin-charge relation in a Landau level, their charge is related to their topological winding number. In addition to being of intrinsic interest, skyrmions can be detected by experimental probes~\cite{Barrett_1995, Schmeller95, Aifer96, Liu22, Coissard22}, and can be used to distinguish ground state phases. 

In this work, we have investigated skyrmion excitations above the $SU(4)$ quantum Hall ferromagnetic ground states at filling factor $\nu=-1$ in monolayer graphene~\cite{Lian_Rosch_Goerbig_2016,Lian_Goerbig_2017}. At this filling, the fourfold nearly-degenerate zeroth Landau level supports a rich set of spin-valley-ordered phases generated by the competition among short-range anisotropic interactions, the Zeeman coupling, and sublattice symmetry-breaking fields. To describe the associated charged topological excitations, we combine an effective field-theory treatment of the long-range Coulomb interaction with a Hartree--Fock description of the symmetry-breaking terms. This approach yields a unified variational energy functional for skyrmion textures, incorporating the spin stiffness, electrostatic self-energy, anisotropy energy, and external-field contributions. By minimizing this functional over the unconstrained variational parameters and the internal $SU(4)$ spinor structure, we determine the energetically favored skyrmion configurations and establish their phase diagram across the anisotropic-interaction parameter space.

In the variational calculation, we parameterize the skyrmion by the spatially varying spinor
\bean
|\chi(\br)\rgl=\sqrt{1-W(r)}\dfrac{x+\text{i}y}{r}|f_1\rgl+\sqrt{W(r)}|f_2\rgl.\nonumber
\eean
The variational degrees of freedom naturally separate into radial and internal components. The radial profile $W(r)$, satisfying the boundary conditions in Eq.~\eqref{eq:W_conditions}, controls the spatial profile of the skyrmion and hence its characteristic size. The spinors $|f_1\rgl \&|f_2\rgl$ determine the orientation of the skyrmion in the internal valley-spin $SU(4)$ space.\\ 
To construct a physically motivated variational ansatz for $W(r)$, we exploit the fact that the nonlinear sigma model (NLSM) contribution $E_\text{stiff}$ is the dominant and $SU(4)$-invariant component of the skyrmion energy. We first determine analytically the radial profile $W_0(r)$ that minimizes $E_\text{stiff}$\cite{Lian_Rosch_Goerbig_2016,Lian_Goerbig_2017,Atteia_Goerbig_2021} (see Appendix~\ref{appdx_1}). Since this solution saturates the Bogomol\'nyi--Prasad--Sommerfield (BPS) bound, we refer to it as the BPS profile. 

A disadvantage of the BPS profile is that it has certain anisotropic contributions to the skyrmion energy that diverge with the system size. To solve this problem, and to give us an additional variational parameter, we deform the BPS profile and generate a one-parameter family of radial profiles
\bean
W_\text{0}(r)=\dfrac{\lambda^2}{r^2+\lambda^2}\quad\rightarrow\quad W_\xi(r)=\bigg(\dfrac{\lambda^2}{\lambda^2+r^2}\bigg)^{1+\xi},\nonumber
\eean
where the deformation parameter $\xi$ accounts for the effects of the Coulomb interaction and the symmetry-breaking terms beyond the NLSM.

By variationally minimizing the skyrmion energy functional atop the four different $\nu=-1$ QHFM ground state phases~\cite{Lian_Rosch_Goerbig_2016, Lian_Goerbig_2017}, we identify thirteen distinct skyrmion states across the anisotropic-interaction parameter space and under different external-field conditions, which are summarized in Table~\ref{table:1}. For some of  these skyrmion configurations, the ground state  spinor $|f_1\rgl$, and the spinor at the skyrmion core  $|f_2\rgl$  can be expressed as direct products of a valley Bloch vector and a spin Bloch vector. In these cases, the skyrmion texture can be readily interpreted in terms of valley flips, spin flips, or simultaneous flips of both. Such skyrmions are characterized by quantized values of the quantity  $\Xi$ defined in Eq.~\eqref{eq:relative_orientation}, which measures the relative orientations of the spin and valley polarizations between $|f_1\rgl$ and $|f_2\rgl$~\cite{Lian_Rosch_Goerbig_2016, Lian_Goerbig_2017}. More generally, however, $|f_1\rgl$ and  $|f_2\rgl$ take spin-valley-entangled forms that cannot be decomposed into direct products of valley and spin states. In this regime, the valley and spin sectors are no longer independent, and the skyrmion texture represents a genuine $SU(4)$ excitation involving coherent spin-valley mixing. We find that this occurs when the sublattice potential $E_V$ is nonzero. In this case, we also find two new skyrmion phases (A-S4 and B-S4) that seem to have been missed in previous works~\cite{Lian_Rosch_Goerbig_2016, Lian_Goerbig_2017}.

We have also examined the robustness of the variational approach with respect to the choice of the radial profile $W(r)$. Besides the continuous parametrization $W_\xi(r)$ adopted in Eq.~\eqref{eq:w(r)}, we considered both its restriction to integer-valued deformation parameters and an alternative ansatz proposed in Refs.~\cite{Lian_Rosch_Goerbig_2016, Lian_Goerbig_2017}. In all cases, the resulting skyrmion energies, optimal spinor configurations, and phase diagrams are in excellent quantitative agreement over the entire parameter space, with energy differences below 1\% and only negligible shifts of the phase boundaries. These results demonstrate that the physical properties of the variationally optimized skyrmions are insensitive to the specific choice of radial ansatz, thereby establishing the robustness and generality of our variational framework.\\
Although the variational framework developed in this work is tailored to skyrmion excitations above the $\nu=-1$ QHFM ground states in monolayer graphene, the overall strategy for constructing the variational energy functional is considerably more general. In particular, it can be naturally extended to the $\nu=0$ quantum Hall regime, where the competition among anisotropic interactions gives rise to a much richer family of spin-valley-entangled QHFM phases~\cite{Das_Kaul_Murthy_2022, De_etal_Murthy2022, Stefanidis_Sodemann2022,an2024uniquely}. The more intricate internal structure of these ground states is expected to support a wider variety of skyrmion textures with novel topological and symmetry properties beyond those identified here.\\
An even more intriguing direction is the extension to fractional QHFM phases in graphene~\cite{Sodemann_MacDonald_2014, Jincheng2024,an2024fractional}. In this case, the long-range Coulomb interaction, strong correlation effects, and internal $SU(4)$ symmetry are expected to compete and cooperate in a highly nontrivial manner, potentially leading to fractional skyrmions with rich internal structures and unconventional properties. The variational framework developed here provides a promising starting point for systematically investigating these topological excitations and their phase diagrams in multicomponent fractional quantum Hall systems~\cite{Park98, Toke07, Balram15, Balram15a}.

\begin{table*}[t]
\centering
\renewcommand{\arraystretch}{1.9} 
\begin{tabular}{|p{2cm}|p{4.3cm}|p{2.8cm}|p{4.3cm}|p{2.3cm}|}
\hline
QHFM  & $\quad|\chi(\infty)\rgl=|f_1\rgl$  & Skyrmions & $\quad|\chi(0)\rgl=|f_2\rgl$  &  $\quad\quad\Xi$ \\ \hhline{|=|=|=|=|=|} 
\multirow{4}{*}{\quad A} & \multirow{4}{*}{$\quad|K,\upa\rgl$} 
      & \quad A-S1 & $\quad|K^\prime,\upa\rgl$ & \quad -2  \\ \cline{3-5}
  &  & \quad A-S2  & $\quad|K,\dwa\rgl$ & \quad 2\\  \cline{3-5}
  &   & \quad A-S3 & $\quad|K^\prime,\dwa\rgl$ & \quad 0\\  \cline{3-5}
  &   & \quad A-S4 \ ($E_V\neq 0$)& $\quad\cos\dfrac{\alpha}{2}|K,\dwa\rgl+\sin\dfrac{\alpha}{2}|K^\prime,\upa\rgl$ & \quad $\pm 2$\\ \hhline{|=|=|=|=|=|} 

  \multirow{4}{*}{\quad B} & \multirow{4}{*}{$\quad|\btau,\upa\rgl$} 
      & \quad B-S1 & $\quad|-\btau,\upa\rgl$ &  \quad -2 \\ \cline{3-5}
  &  & \quad B-S2  & $\quad|\btau,\dwa\rgl$ & \quad 2 \\  \cline{3-5}
  &   & \quad B-S3 & $\quad|-\btau^\prime,\dwa\rgl$ & Unquantized \\  \cline{3-5}
  &   & \quad B-S4 \ ($E_V\neq 0$)& $\quad\cos\dfrac{\alpha}{2}|-\btau,\upa\rgl+\sin\dfrac{\alpha}{2}|\btau^\prime,\dwa\rgl$ & Unquantized\\ \hhline{|=|=|=|=|=|} 

  \multirow{3}{*}{\quad C} & \multirow{3}{*}{$\quad\cos\dfrac{\alpha}{2}|\btau,\upa\rgl-\sin\dfrac{\alpha}{2}|-\btau,\dwa\rgl$} 
      & \quad C-S \ ($E_V\neq 0$)& \quad Eq.~\eqref{eq:C-S_spinor} &  Unquantized \\ \cline{3-5}
  &  & \quad C-S1  \ ($E_V= 0$)& $\quad\sin\dfrac{\alpha}{2}|\hat e_x,\upa\rgl+\cos\dfrac{\alpha}{2}|-\hat e_x,\dwa\rgl$ & \quad 0\\  \cline{3-5}
  &   & \quad C-S2 \ ($E_V= 0$)& $\quad\sin\dfrac{\alpha^\prime}{2}|-\hat e_x,\upa\rgl+\cos\dfrac{\alpha^\prime}{2}|\hat e_x,\dwa\rgl$ & \quad -2\\ \hhline{|=|=|=|=|=|} 

  \multirow{2}{*}{\quad D} & \multirow{2}{*}{$\quad\cos\dfrac{\alpha}{2}|K,\upa\rgl+\sin\dfrac{\alpha}{2}|K^\prime,\dwa\rgl$} 
      & \quad D-S1 & $\quad\cos\dfrac{\alpha^\prime}{2}|K,\dwa\rgl+\sin\dfrac{\alpha^\prime}{2}|K^\prime,\upa\rgl$ &  \quad -2 \\ \cline{3-5}
  &  & \quad D-S2  & $\quad \sin\dfrac{\alpha}{2}|K,\upa\rgl-\cos\dfrac{\alpha}{2}|K^\prime,\dwa\rgl$ & \quad 0\\  \cline{3-5}
\hline 
\end{tabular}
\captionsetup{justification=justified, singlelinecheck=false, width=0.95\textwidth}
\caption{Skyrmion states atop all possible $\nu=-1$ QHFM ground state phases. In total, thirteen distinct skyrmion states are identified. Among them, A-S4, B-S4, and C-S exist only in the presence of a finite sublattice symmetry-breaking potential $E_V$. Furthermore, C-S splits into two distinct skyrmion states, C-S1 and C-S2, in the limit of $E_V\rightarrow0$. The corresponding values of the discriminator $\Xi$ defined in Eq.~\eqref{eq:relative_orientation} are also listed. The label "Unquantized" indicates that $\Xi$ takes continuous, non-integer values.  }\label{table:1}
\end{table*}

\section{Acknowledgements}
J.A. acknowledges support from the Shandong Province Natural Science Foundation (Grant No. ZR2026QC0001). J.A. is also grateful to the University of Kentucky Center for Computational Sciences and Information Technology Services Research Computing for their support and use of the Lipscomb Compute Cluster and associated research computing resources. ACB acknowledges financial support from the Anusandhan National Research Foundation (ANRF) of the Department of Science and Technology (DST) via the Mathematical Research Impact Centric Support (MATRICS) Grant No. MTR/2023/000002 and the Advanced Research Grant No. ANRF/ARG/2025/000562/PS. GM is grateful to the United States Department of Energy for partial support under DE-SC0024346, and to Indiana University, the Pennsylvania State University, and ICTS-TIFR Bangalore for their hospitality.

\bibliography{hall}

\begin{appendices}

\setcounter{figure}{0}
\renewcommand{\thefigure}{A.\arabic{figure}}
\setcounter{equation}{0}
\renewcommand{\theequation}{A\arabic{equation}}

\section{Minimization of $E_\text{stiff}$}\label{appdx_1}

Recall that the total variational energy of a skyrmion configuration has three parts,
\begin{equation}
    E_\text{skyr}=E_\text{stiff}+E_\text{Coul}+E_\text{an},
\end{equation}
the nonlinear sigma model stiffness energy $E_\text{stiff}$, the Coulomb energy $E_\text{C}$, and the energy of the $SU(4)$ symmetry-breaking parts of the Hamiltonian $E_\text{an}$, which contain the one-body fields as well as the anisotropic short-range interactions. 
Typically, $E_\text{stiff}$ provides the leading contribution to the total skyrmion energy. Following previous work, we first find the optimal skyrmion configuration that minimizes $E_\text{stiff}$~\cite{Goerbig_nu0_Skyrmion_Zoo_2021}, and subsequently consider deformations away from this optimal configuration to minimize other terms. Since the projector to the occupied state (also the order parameter) $\bP(\br)$ is positive semi-definite, we construct the following operator
\bean
\bA(\br)=\sqrt{\bP(\br)}\partial\bP(\br),\ \Rightarrow\ \bA^\dagger(\br)=\bar\partial\bP(\br)\sqrt{\bP(\br)},
\eean 
where $\partial=\partial_x-i\partial_y,\ \bar\partial=\partial_x+i\partial_y$. By construction, one obtains the inequality
\bean
\label{eq: inequality}
\text{Tr}\big[\bA^\dagger\bA\big]=\text{Tr}\big[\bar\partial\bP\bP\partial\bP\big]\geq0.
\eean 
By virtue of the relation $\bP=\bP^2$, valid for $\nu=-1$, we have 
\bean
\partial_\mu\bP=\partial_\mu\bP^2=\bP\partial_\mu\bP+\big(\partial_\mu\bP\big)\bP.
\eean 
Then the inequality stated in Eq.~\eqref{eq: inequality} can be explicitly expressed as 
\bean
\text{Tr}\big[\partial_\mu\bP\partial_\mu\bP\big]+2i\varepsilon_{\mu\nu}\text{Tr}\big[\bP\partial_\mu\bP\partial_\nu\bP\big]\geq0.
\eean 
Upon integration over the two-dimensional space, this inequality yields the NLSM energy, $E_\text{stiff}$ of Eq.~\eqref{eq:NLSM}, and the associated topological charge given in Eqs.~\eqref{eq:topological_charge_density}-\eqref{eq:topo_charge}, leading to the celebrated Bogomol\'nyi–Prasad–Sommerfield (BPS) bound~\cite{Goerbig_nu0_Skyrmion_Zoo_2021, Lian_Goerbig_2017, Macfarlane1979, Bogomolny:1975de, Prasad_Sommerfield_1975}, 
\bean
E_\text{stiff}\geq 4\pi\rho_s|Q|.
\eean 
$E_\text{stiff}$ reaches its minimum, $E_\text{stiff}= 4\pi\rho_s|Q|$ when 
\bean
\label{eq: BPS_bound_saturation_P}
\bP\partial\bP=0\ \Leftrightarrow\ \bar\partial\bP\bP=0.
\eean 
The simplest skyrmionic solution with $Q=1$ to Eq.~\eqref{eq: BPS_bound_saturation_P} is given by~\cite{Goerbig_nu0_Skyrmion_Zoo_2021, lian_PhD_thesis_2017} 
\bean\label{eq:spinor_chi}
P^S(\br)=|\chi(\br)\rgl\lgl \chi(\br)|,\quad |\chi(\br)\rgl=\dfrac{z|f_1\rgl+\lambda|f_2\rgl}{\sqrt{r^2+\lambda^2}},
\eean 
with $\lambda$ being the characteristic length of the skyrmion.  The function $W(r)$ defined in Eq.~\eqref{eq:P(r)}  for the BPS state/solution is
\bean
W_0(r)=\dfrac{\lambda^2}{r^2+\lambda^2}.
\eean

\section{Evaluation of Symmetry-Breaking Terms}
\label{appdx_2}
In this Appendix, we evaluate the contributions to the skyrmion excited energy from the symmetry-breaking terms. When performing the Hartree-Fock approximation for the two-body anisotropic interacting terms in Eq.~\eqref{eq: anisotropic_interaction} with an SSD state $|\Psi\rgl$, we decompose the average of four-fermi terms as the product of two-fermi averages:
\bean\label{eq:app_HF_approx}
\hspace{-0.5cm}&&\lgl\Psi|\hat H_\text{an}|\Psi\rgl\nn
\hspace{-0.5cm}&=&\dfrac{1}{2}\sum_{a=x,y,z}\int d^2\br_1d^2\br_2V_a(\br_1-\br_2)\tau^a_{\alpha\gamma}\tau^a_{\beta\eta}\nn
\hspace{-0.5cm}&&\times\bigg[\lgl\Psi|\hpsid_\alpha(\br_1)\hpsi_\gamma(\br_1)|\Psi\rgl\lgl\Psi|\hpsid_\beta(\br_2)\hpsi_\eta(\br_2)|\Psi\rgl,\nn
\hspace{-0.5cm}&&\quad -\lgl\Psi|\hpsid_\alpha(\br_1)\hpsi_\eta(\br_2)|\Psi\rgl\lgl\Psi|\hpsid_\beta(\br_2)\hpsi_\gamma(\br_1)|\Psi\rgl\bigg].
\eean
To carry this out, one needs to evaluate the general one-body density matrix in the resulting Fock terms, 
\bean
&&\quad\lgl\Psi|\hpsid_\alpha(\br_1)\hpsi_{\eta}(\br_2)|\Psi\rgl\nn
&&=\sum_{m,m^\prime}\phi^*_{m}(\br_1)\phi_{m^\prime}(\br_2)\underbrace{\lgl\Psi|\hcd_{m,\alpha}\hc_{m^\prime,\eta}|\Psi\rgl}_{\equiv \cM_{m,\alpha;m^\prime,\eta}}.
\eean
To proceed, we construct a one-body operator in the LLL using the matrix elements $\cM$ defined above
\bean
\hat\cM_{\alpha\eta}=\sum_{m,m^\prime}|m\rgl \cM_{m,\alpha;m^\prime,\eta}\lgl m^\prime|.
\eean 
Thus, we have 
\bean
\lgl\Psi|\hpsid_\alpha(\br_1)\hpsi_{\eta}(\br_2)|\Psi\rgl=\lgl\br_1|\hat\cM_{\alpha\eta}|\br_2\rgl.
\eean 
Let us introduce the LLL coherent state~\cite{Girvin_Jach1984_PRB},
\bean
|\bR\rgl=\sum_m\sqrt{2\pi\ell^2}\phi^*_m(\bR)|m\rgl.
\eean 
Using this, one can construct the LLL projector,
\bean
\hat\Pi=\int \dfrac{d^2\bR}{2\pi\ell^2}|\bR\rgl\lgl\bR|,
\eean 
the matrix elements of which are given by 
\bean
\lgl\br_1|\hat\Pi|\br_2\rgl&=&
\int\dfrac{d^2\bR}{2\pi\ell^2}\lgl\br_1|\bR\rgl\lgl\bR|\br_2\rgl\nn
&=&\sum_{mm^\prime}\int d^2\bR\phi_m(\br_1)\phi^*_m(\bR)\phi_{m^\prime}(\bR)\phi^*_{m^\prime}(\br_2)\nn
&=&\sum_m\phi_m(\br_1)\phi^*_m(\br_2)\nn
&=&\dfrac{1}{2\pi\ell^2}\text{Exp}\Big[-\dfrac{|\br_1-\br_2|^2}{4\ell^2}+i\dfrac{(\br_1\times\br_2)_z}{2\ell^2}\Big]\nn
&\equiv&\Pi(\br_1,\br_2).
\eean 
Accordingly, the wave function of the coherent state can be expressed as  
\bean
\hspace{-0.3cm}\lgl\br|\bR\rgl=\sqrt{2\pi\ell^2}\sum_m\phi_m(\br)\phi^*_m(\bR)=\sqrt{2\pi\ell^2}\Pi(\br,\bR).
\eean 
Since the operator $\hat \cM$ acts within the LLL, we have 
\bean
\hat\cM=\hat\Pi\hat\cM\hat\Pi,
\eean 
or more explicitly, 
\bean
\hspace{-0.5cm}&&\lgl\br_1|\hat\cM|\br_2\rgl\nn
\hspace{-0.5cm}&=&\int \dfrac{d^2\bR}{2\pi\ell^2}\dfrac{d^2\bR^\prime}{2\pi\ell^2}\lgl \br_1|\bR\rgl\lgl\bR|\hat\cM|\bR^\prime\rgl\lgl\bR^\prime|\nn
\hspace{-0.5cm}&=&\int d^2\bR d^2\bR^\prime\Pi(\br_1,\bR)\dfrac{\lgl\bR|\hat\cM|\bR^\prime\rgl}{2\pi\ell^2}\Pi(\bR^\prime,\br_2).
\eean
Given that the interactions in Eq.~\eqref{eq:app_HF_approx} are short ranged, namely, $|\br_1-\br_2|\approx\ell$ and the LLL kernels $\Pi(\br_1,\bR)\ \&\ \Pi(\bR^\prime,\br_2)$ are localized within approximately a magnetic length around $\br_1$ and $\br_2$ respectively, the double convolution is dominated by $\bR\approx\bR^\prime$. We can therefore ignore the spatial dependence of the matrix elements of $\cM$ in the coherent state basis, i.e.
\bean
\lgl\br_1|\hat\cM|\br_2\rgl=\int d^2\bR\Pi(\br_1,\bR)\bP(\bR)\Pi(\bR,\br_2),
\eean 
where 
\bean
\bP(\bR)=\lgl\bR|\hat\cM|\bR\rgl
\eean
is the local order parameter. We are assuming that $\bP$ varies very slowly on the scale of the magnetic length when it comes to evaluating the anisotropic energy, that is, $\ell\ll\lambda$. The terms we drop will come with derivatives of $\bP$, and in principle will correct the stiffness energy, but since the anisotropic couplings are small, we ignore these corrections. Since first derivatives in space cannot occur due to rotational invariance, we are assuming that 
\begin{equation}
u_a \left(\frac{\ell}{\lambda}\right)^2\ll \rho_s,
\end{equation}
a condition easily satisfied for realistic samples.  

Now we go to the center-of-mass(COM) and relative coordinates,
\bean
\bR_c=\dfrac{\br_1+\br_2}{2},\ \bs=\br_1-\br_2,
\eean 
whence we obtain 
\bean
\hspace{-0.8cm}&&\lgl\Psi|\hpsid_{\alpha}(\br_1)\hpsi_\eta(\br_2)|\Psi\rgl\nn
\hspace{-0.8cm}&=&\int d^2\bR\Pi(\bR_c+\dfrac{\bs}{2},\bR)\bP_{\eta\alpha}(\bR)\Pi(\bR,\bR_c-\dfrac{\bs}{2})\nn
\hspace{-0.8cm}&=&\dfrac{1}{(2\pi\ell^2)^2}\text{Exp}\Big[-\dfrac{|\bs|^2}{8\ell^2}+i\dfrac{(\bs\times\bR_c)_z}{2\ell^2}\Big]\nn
\hspace{-0.8cm}&&\times\int d^2\bu \bP_{\eta\alpha}(\bR_c+\bu)\text{Exp}\Big[-\dfrac{|\bu|^2}{2\ell^2}+i\dfrac{(\bs\times\bu)_z}{2\ell^2}\Big].\nonumber
\eean 
Note that we have made the replacement $\bu\equiv\bR-\bR_c$.
As mentioned above, assuming $\ell\ll\lambda$, we approximate  $\bP_{\eta\alpha}(\bR_c+\bu)=\bP_{\eta\alpha}(\bR_c)$, which leads to 
\bean\label{eq:app_general_density_matrix}
\hspace{-0.8cm}&&\quad\lgl\Psi|\hpsid_\alpha(\br_1)\hpsi_{\eta}(\br_2)|\Psi\rgl\nn
\hspace{-0.8cm}&&=\dfrac{1}{2\pi\ell^2}\bP_{\eta\alpha}(\bR_c)\text{Exp}\bigg[{-\dfrac{|\bs|^2}{4\ell^2}}+i\dfrac{(\bs\times\bR_c)_z}{2\ell^2}\bigg].
\eean 
The point of this exercise is to express the expectation value $\lgl\Psi|\hat H_\text{an}|\Psi\rgl$ as a functional of the local order parameter $\bP$. \\
More explicitly, given a skyrmion state with a local order parameter
\bean
\lgl \Psi_s|\hpsid_\alpha(\br)\hpsi_\gamma(\br)|\Psi_s\rgl=\dfrac{1}{2\pi\ell^2}P^S_{\gamma\alpha}(\br),
\eean  
the expectation value of the anisotropic interaction Hamiltonian is
\bea
\hspace{-0.4cm}&&E_\text{an}=\lgl \Psi_s|\hat H_\text{an}|\Psi_s\rgl\nn
\hspace{-0.4cm}&&\quad\quad=\dfrac{1}{2}\int d^2\br_1d^2\br_2V_a(\br_1-\br_2)\tau^a_{\alpha\gamma}\tau^a_{\beta\eta}\Big[\dfrac{P^S_{\gamma\alpha}(\br_1)}{2\pi\ell^2}\dfrac{P^S_{\eta\beta}(\br_2)}{2\pi\ell^2}\nn
\hspace{-0.4cm}&&\quad\quad\quad\ -\lgl \Psi_s|\hpsid_\alpha(\br_1)\hpsi_\eta(\br_2)|\Psi_s\rgl\lgl \Psi_s|\hpsid_\beta(\br_2)\hpsi_\gamma(\br_1)|\Psi_s\rgl\Big].
\eea 
In the Hartree term above, using the fact that the anisotropic interactions are nonzero only up to a few $\ell$ and that $\ell\ll\lambda$, we can replace the density-density interacting term $P^S(\br_1)P^S(\br_2)$ by $P^S(\bR_c)^2$. Similarly, in the Fock term, we use Eq.~\eqref{eq:app_general_density_matrix} for $|\Psi_s\rgl$. Consequently,
\onecolumngrid
\bean\label{eq:app_ess}
\hspace{-1cm}&&E_\text{an}=
\dfrac{1}{2}\int \dfrac{d^2\bs}{2\pi\ell^2} V_a(\bs)\int \dfrac{d^2\bR_c}{2\pi\ell^2}\text{Tr}\Big[P^S(\bR_c)\tau^a\Big]^2-\dfrac{1}{2}\int \dfrac{d^2\bs}{2\pi\ell^2} V_a(\bs)e^{-\frac{|\bs|^2}{2\ell^2}}\int \dfrac{d^2\bR_c}{2\pi\ell^2} \text{Tr}\Big[P^S(\bR_c)\tau^aP^S(\bR_c)\tau^a\Big]\nn
\hspace{-1cm}&&\quad\quad=\dfrac{1}{2}u_{a,H}\int \dfrac{d^2\bR_c}{2\pi\ell^2}\Big[\big(P^S_{11}\tau^a_{11}\big)^2+2P^S_{12}\tau^a_{21}P^S_{21}\tau^a_{12}+2P^S_{11}\tau^a_{11}P^S_{22}\tau^a_{22}+\big(P^S_{22}\tau^a_{22}\big)^2\Big]\nn
\hspace{-1cm}&&\quad\quad\ -\dfrac{1}{2}u_{a,F}\int \dfrac{d^2\bR_c}{2\pi\ell^2}\Big[\big(P^S_{11}\tau^a_{11}\big)^2+2P^S_{12}\tau^a_{22}P^S_{21}\tau^a_{11}+2P^S_{11}\tau^a_{12}P^S_{22}\tau^a_{21}+\big(P^S_{22}\tau^a_{22}\big)^2\Big]\nn
\hspace{-1cm}&&\quad\quad=\dfrac{1}{2}\big(u_{a,H}-u_{a,F}\big)\int \dfrac{d^2\bR_c}{2\pi\ell^2}\bigg[(\tau^a_{11})^2\big(1-W(R_c)\big)^2+2(\tau^a_{11}\tau^a_{22}+\tau^a_{12}\tau^a_{21})W(R_c)\big(1-W(R_c)\big)+(\tau^a_{22})^2W^2(R_c)\bigg]\nn
\hspace{-1cm}&&\quad\quad=\dfrac{1}{2}\big(u_{a,H}-u_{a,F}\big)\big(\tau^a_{11}\big)^2\dfrac{\pi\Lambda^2}{2\pi\ell^2}+\big(u_{a,H}-u_{a,F}\big)\Big[\tau^a_{11}\tau^a_{22}+\tau^a_{12}\tau^a_{21}-\big(\tau^a_{11}\big)^2\Big]\Big(\frac{\lambda}{\ell}\Big)^2Y_1(\xi)\nn
\hspace{-1cm}&&\quad\quad\quad\quad\quad\quad\quad\quad\quad\quad\quad\quad\quad\quad\quad+\dfrac{1}{2}\big(u_{a,H}-u_{a,F}\big)\Big[\big(\tau^a_{11}-\tau^a_{22}\big)^2-2\tau^a_{12}\tau^a_{21}\Big]\Big(\frac{\lambda}{\ell}\Big)^2Y_2(\xi),
\eean 
\twocolumngrid
where we have introduced a cutoff for the divergent integral $\int\frac{d^2\br}{2\pi\ell^2}=\frac{\pi\Lambda^2}{2\pi\ell^2}\equiv\frac{A}{2\pi\ell^2}=N_\phi$, where $A$ is the area of the system. We have also defined
\bean
\hspace{-1cm}&&\Big(\frac{\lambda}{\ell}\Big)^2Y_1(\xi)=\int \dfrac{d^2\br}{2\pi\ell^2}W(r)=\dfrac{\lambda^2}{\ell^2}\dfrac{1}{2\xi},\label{eq:app_integral_over_W}\\
\hspace{-1cm}&&\Big(\frac{\lambda}{\ell}\Big)^2Y_2(\xi)=\int \dfrac{d^2\br}{2\pi\ell^2}W^2(r)=\dfrac{\lambda^2}{\ell^2}\dfrac{1}{2(1+2\xi)}.\label{eq:app_integral_over_W_sqrt}
\eean 
In the first equality of both the equations above, $W(r)$ is generic, whereas in the second equality we assume it is the $W_\xi(r)$ defined in Eq.~\eqref{eq:w(r)}.\\
Recall that the Hartree and Fock couplings  are given by 
\bean
\hspace{-1cm}&&u_{a,H}\equiv\int\dfrac{d^2\bs}{2\pi\ell^2}  V_a(\bs)=\dfrac{v_a(0)}{2\pi\ell^2},\nn
\hspace{-1cm}&&u_{a,F}\equiv\int \dfrac{d^2\bs}{2\pi\ell^2}  V_a(\bs)e^{-\frac{|\bs|^2}{2\ell^2}}=\int\dfrac{d^2\bq}{(2\pi)^2}v_a(\bq)e^{-\frac{q^2\ell^2}{2}},
\eean 
which  agrees with Eq.~\eqref{eq:HF_couplings} for the explicit form of $v_a(\bq)$ in Eq.~\eqref{eq:vaq_def}. As in the ground-state energy functional, Eq.~\eqref{eq:_m1_ground_state_energy}, the anisotropic-interaction contribution to the skyrmion energy functional depends only on the difference between the Hartree and Fock couplings, $(u_{a, H}-u_{a, F})$, indicating the interactions in the even angular momentum channels  do not contribute to the skyrmion energy~\cite{QHFM_Moon_etal_1995,Shivaji_Skyrmion,Fertig_skyrmion_1994,Fertig_skyrmion_1996}. In particular, the USR contact interaction does not contribute. 
The 1-body Hamiltonian representing the external fields takes the following form in real space
\bean
\hspace{-1cm}&&\hat H_Z=-E_Z\int{d^2\br}\hpsid_\alpha(\br)\big(\tau^0_\text{valley}\otimes\sigma^z_\text{spin}\big)_{\alpha\gamma}\hpsi_\gamma(\br),\nn
\hspace{-1cm}&&\hat H_V=-E_V\int{d^2\br}\hpsid_\alpha(\br)\big(\tau^z_\text{valley}\otimes\sigma^0_\text{spin}\big)_{\alpha\gamma}\hpsi_\gamma(\br).
\eean 
The Zeeman energy of the skyrmion is
\bean\label{eq:app_zeeman}
\hspace{-1cm}&&\quad\lgl\Psi_s|\hat H_Z|\Psi_s\rgl\nn
\hspace{-1cm}&&=-E_Z\int\dfrac{d^2\br}{2\pi\ell^2}\Big[P^S_{11}(\br)\sigma^z_{11}+P^S_{22}(\br)\sigma^z_{22}\Big].
\eean
Using the explicit ansatz of Eq.~\eqref{eq:w(r)}  for $W_\xi(r)$, we get  
\bean\label{eq:app_zeeman_w0}
\hspace{-1cm}&&\quad\lgl\Psi_s|\hat H_Z|\Psi_s\rgl\nn
\hspace{-1cm}&&=-E_Z\sigma^z_{11}\dfrac{\pi\Lambda^2}{2\pi\ell^2}+E_Z\big(\sigma^z_{11}-\sigma^z_{22}\big)\Big(\frac{\lambda}{\ell}\Big)^2Y_1(\xi),
\eean 
where $Y_1(\xi)$ is given in Eq.~\eqref{eq:app_integral_over_W}.\\

Similarly, the valley Zeeman energy is
\bean\label{eq:app_valley_zeeman}
\hspace{-0.6cm}&&\lgl\Psi_s|\hat H_V|\Psi_s\rgl=-E_V\tau^z_{11}\dfrac{\pi\Lambda^2}{2\pi\ell^2}+E_V\big(\tau^z_{11}-\tau^z_{22}\big)\Big(\frac{\lambda}{\ell}\Big)^2Y_1(\xi).\nn
\hspace{-0.6cm}&&\quad 
\eean 
The expressions in Eq.~\eqref{eq:app_zeeman} and \eqref{eq:app_valley_zeeman} are valid for generic $W(r)$ as indicated in Eq.~\eqref{eq:integrals_general_W}.

After collecting Eqs.~\eqref{eq:app_ess}, \eqref{eq:app_zeeman} and \eqref{eq:app_valley_zeeman}, we arrive at the expression for the evaluation of all $SU(4)$ symmetry-breaking terms, 
\bean\label{eq:app_symmetry_breaking_energy}
\hspace{-1cm}&&E_\text{SB}=\lgl\Psi_s|\hat H_\text{an}+\hat H_Z+\hat H_V|\Psi_s\rgl\nn
\hspace{-1cm}&&\quad\quad =\dfrac{\pi\Lambda^2}{2\pi\ell^2}E_\text{quad}+\dfrac{\lambda^2}{\ell^2}\bigg[\dfrac{\cE_1}{2\xi}+\dfrac{\cE_2}{2(1+2\xi)}\bigg],
\eean 
where 
\bean
\hspace{-0.6cm}E_\text{quad}=\dfrac{1}{2}\big(u_{a,H}-u_{a,F}\big)\big(\tau^a_{11}\big)^2-E_Z\sigma^z_{11}-E_V\tau^z_{11},
\eean
which is exactly $E_\text{g.s.}$ in Eq.~\eqref{eq:_m1_ground_state_energy} and 
\bean
\hspace{-1cm}&&\cE_1=\big(u_{a,H}-u_{a,F}\big)\Big[\tau^a_{12}\tau^a_{21}+\tau^a_{11}\tau^a_{22}-\big(\tau^a_{11}\big)^2\Big]\nn
\hspace{-1cm}&&\quad\quad\ +E_Z\big(\sigma^z_{11}-\sigma^z_{22}\big)+E_V\big(\tau^z_{11}-\tau^z_{22}\big),\label{eq:app_E1}\\
\hspace{-1cm}&&\cE_2=\dfrac{1}{2}\big(u_{a,H}-u_{a,F}\big)\Big[\big(\tau^a_{11}-\tau^a_{22}\big)^2-2\tau^a_{12}\tau^a_{21}\Big]\label{eq:app_E2},
\eean 
which are functionals of the spinor in the skyrmion core $|\chi(0)\rgl=|f_2\rgl$.

\end{appendices}

\end{document}